\documentclass[journal]{IEEEtran}
\IEEEoverridecommandlockouts
\usepackage{cite}
\usepackage{amsmath,amssymb,amsfonts}
\usepackage{graphicx}
\usepackage{textcomp}
\usepackage{xcolor}
\usepackage{algorithm}
\usepackage{algorithmic}
\usepackage{enumitem}
\usepackage{footnote}
\usepackage{multirow}
\usepackage{threeparttable}
\usepackage{makecell}
\usepackage[colorlinks,linkcolor=blue]{hyperref}
\ifCLASSOPTIONcompsoc
\usepackage[caption=false, font=normalsize, labelfont=sf, textfont=sf]{subfig}
\else
\usepackage[caption=false, font=footnotesize]{subfig}
\fi

\usepackage{array}

\def\BibTeX{{\rm B\kern-.05em{\sc i\kern-.025em b}\kern-.08em
    T\kern-.1667em\lower.7ex\hbox{E}\kern-.125emX}}

\makeatletter
\renewcommand{\maketag@@@}[1]{\hbox{\m@th\normalsize\normalfont#1}}%
\makeatother

\begin{document}

\title{Low-Overhead Channel Estimation via 3D Extrapolation for TDD mmWave Massive MIMO Systems Under High-Mobility Scenarios}
\author{Binggui Zhou,
        Xi Yang,
        Shaodan Ma,
        Feifei Gao,
        and Guanghua Yang\vspace{-2em}
\thanks{This work was supported in part by the National Natural Science Foundation of China under Grants 62261160650, 62171201, and 62301221; in part by the Major Talent Program of Guangdong Provincial under Grant 2019QN01S103; in part by the open research fund of National Mobile Communications Research Laboratory, Southeast University (No. 2024D02); in part by the Science and Technology Development Fund, Macau SAR, under Grants 0087/2022/AFJ and 001/2024/SKL; and in part by the Research Committee of University of Macau under Grants MYRG-GRG2023-00116-FST-UMDF and MYRG2020-00095-FST. (Corresponding authors: Xi Yang; Guanghua Yang.)}
\thanks{Binggui Zhou is with the School of Intelligent Systems Science and Engineering, Jinan University, Zhuhai 519070, China; and also with the State Key Laboratory of Internet of Things for Smart City and the Department of Electrical and Computer Engineering, University of Macau, Macao 999078, China (e-mail: binggui.zhou@connect.um.edu.mo).}
\thanks{Xi Yang is with the Shanghai Key Laboratory of Multidimensional Information Processing, East China Normal University, Shanghai 200241, China, and is also with the National Mobile Communications Research Laboratory, Southeast University, Nanjing 210096, China (e-mail: xyang@cee.ecnu.edu.cn).}
\thanks{Shaodan Ma is with the State Key Laboratory of Internet of Things for Smart City and the Department of Electrical and Computer Engineering, University of Macau, Macao 999078, China (e-mail: shaodanma@um.edu.mo).}
\thanks{Feifei Gao is with the Department of Automation, Tsinghua University, Beijing 100084, China (e-mail: feifeigao@ieee.org).}
\thanks{Guanghua Yang is with the School of Intelligent Systems Science and Engineering, and Guangdong International Cooperation Base of Science and Technology for GBA Smart Logistics, Jinan University, Zhuhai 519070, China (e-mail: ghyang@jnu.edu.cn).}
}

\maketitle

\begin{abstract}
In time division duplexing (TDD) millimeter wave (mmWave) massive multiple-input multiple-output (MIMO) systems, downlink channel state information (CSI) can be obtained from uplink channel estimation thanks to channel reciprocity. However, under high-mobility scenarios, frequent uplink channel estimation is needed due to channel aging. Additionally, large amounts of antennas and subcarriers result in high-dimensional CSI matrices, aggravating pilot training overhead. To address this, we propose a three-domain (3D) channel extrapolation framework across spatial, frequency, and temporal domains. First, considering the effectiveness of traditional knowledge-driven channel estimation methods and the marginal effects of pilots in the spatial and frequency domains, a knowledge-and-data driven spatial-frequency channel extrapolation network (KDD-SFCEN) is proposed for uplink channel estimation via joint spatial-frequency channel extrapolation to reduce spatial-frequency domain pilot overhead. Then, leveraging channel reciprocity and temporal dependencies, we propose a temporal uplink-downlink channel extrapolation network (TUDCEN) powered by generative artificial intelligence for slot-level channel extrapolation, aiming to reduce the tremendous temporal domain pilot overhead caused by high mobility. Numerical results demonstrate the superiority of the proposed framework in significantly reducing the pilot training overhead by $16$ times and improving the system's spectral efficiency under high-mobility scenarios compared with state-of-the-art channel estimation/extrapolation methods.
\end{abstract}


\section{Introduction}

\IEEEPARstart MILLIMETER wave (mmWave) massive multi-input multi-output (MIMO) has been recognized as a pivotal technology for the fifth generation (5G) wireless communication systems and beyond \cite{larsson2014massive, feng2019twoway, feng2018impact}. mmWave massive MIMO is anticipated to provide spatial degrees of freedom, diversity or multiplexing gain, and array gain, thereby improving the spectral and energy efficiencies of wireless communication systems. To reap the benefits of massive MIMO, accurate channel state information (CSI) should be attained whether operating in the time division duplexing (TDD) or frequency division duplexing (FDD) modes\cite{zhou2024lowoverhead,chen2022highaccuracy}. Specifically, in TDD massive MIMO systems, thanks to the uplink-downlink channel reciprocity, the base station (BS) can obtain the downlink CSI via uplink channel estimation and downlink CSI derivation at the BS side.

The uplink-downlink channel reciprocity in the TDD massive MIMO systems holds within the channel coherence time. However, in high-mobility scenarios, the channel is fast time-varying due to the UE movement, resulting in a short channel coherence time and the consequent channel aging issue, i.e., the channel varies between when it is acquired at the BS and when it is used for downlink precoding\cite{truong2013effects}. In addition, compared with sub-6 GHz bands, mmWave bands are more vulnerable to the mobility of the user equipment (UE) since higher frequency bands generally lead to a shorter channel coherence time\cite{marzetta2016fundamentals}, deteriorating the channel aging issue in mmWave massive MIMO systems under high-mobility scenarios. To avoid significant performance degradation caused by channel aging, frequent channel estimation needs to be conducted such that downlink precoding can be conducted based on up-to-date CSI, leading to huge temporal domain pilot training overhead. Besides, due to the large number of antennas at the BS and large amounts of subcarriers in orthogonal frequency-division multiplexing (OFDM) systems, the spatial and frequency domain pilot training overhead is also huge and unacceptable. In addition,  to reduce the hardware cost and power consumption, a hybrid precoding structure is usually adopted and only a small number of radio frequency (RF) chains are deployed at the BS, especially for mmWave massive MIMO systems. To obtain the uplink CSI at all receiving antennas at the BS, the BS has to switch the antennas connected to these RF chains several times during the uplink channel estimation, leading to substantial time and power consumption. Therefore, it is of great significance to minimize the pilot training overhead in the spatial, frequency, and temporal domains while estimating the CSI accurately.

To circumvent huge frequency domain pilot training overhead, frequency domain channel extrapolation has been widely investigated in related works. In \cite{lu2009least}, a linear interpolation least square (LS) channel estimation method was proposed for channel estimation with pilots in partial subcarriers. In \cite{kusaykin2021based},  the performance of various interpolation methods for mmWave MIMO-OFDM systems, including spline interpolation, discrete Fourier transform (DFT) -based interpolation, etc., was investigated. Recently, with the advances in deep learning, many deep learning-based frequency domain channel extrapolation methods were proposed to further reduce pilot training overhead and improve channel estimation accuracy from the frequency domain perspective. In \cite{soltani2019deep}, the super-resolution convolutional neural network (SRCNN) was presented to obtain full channel responses with the channel responses at pilot positions via deep image processing techniques. By jointly learning the spatial-temporal domain features of massive MIMO channels with a temporal attention module and a spatial attention module, a dual-attention-based channel estimation network (DACEN) was proposed to realize accurate channel estimation via low-density pilots in the frequency domain\cite{zhou2024pay}.

The development of deep learning has also led to increasing attention in antenna (spatial) domain channel extrapolation in recent years, aiming at reducing the huge spatial domain pilot training overhead and prohibitive time and power consumption. For example, two fully connected neural networks (FCNNs) were proposed to use the CSI of a subset of antennas to extrapolate the CSI of other antennas\cite{lin2021deepa, yang2020deepa}.

In addition to spatial and frequency domain channel extrapolation, channel prediction (i.e., temporal channel extrapolation) is also a convincing method for alleviating the huge pilot training overhead in high-mobility scenarios. Linear extrapolation methods \cite{bui2013performance} and statistical prediction models, e.g., autoregressive (AR) models \cite{baddour2005autoregressivea, duel-hallen2007fading}, were proposed and have demonstrated that channel prediction is promising to mitigate the impacts of channel aging. The advancements of DNNs further improve the channel prediction performance and thus would further reduce the channel estimation frequency in high-mobility scenarios. In \cite{mattu2022deep}, a novel long short-term memory (LSTM) based channel predictor was proposed to learn channel variations and thereby reduce the pilot training overhead for channel estimation. An attention-based channel predictor was proposed in \cite{jiang2022accurate} to achieve frame-level channel estimates for mobile scenarios. However, frame-level channel prediction methods are unable to achieve high spectral efficiency due to fast time-varying channels under high-mobility scenarios, posing demands on slot-level channel extrapolation. Moreover, existing channel prediction methods are unable to deal with massive MIMO-OFDM systems with substantial antennas and subcarriers due to the high-dimensional CSI matrices and the consequent tremendous computational complexity.

To overcome the limitations of these existing works and to systematically reduce the pilot overhead, a spatial, frequency, and temporal domain (3D) channel extrapolation framework is proposed in this paper to reduce the pilot training overhead from these three domains respectively. First, it can be observed that the number of pilots in one certain domain shows the marginal effect, i.e., as the number of pilots in this domain increases, the improvement in channel estimation accuracy gradually decreases.\footnote{This observation is further validated with our simulation results in Section \ref{Sec. Sim. B}.} Second, considering the capability of deep neural networks (DNNs) in extracting the spatial and frequency domain characteristics of massive MIMO channels, it is expected that exploiting DNNs for joint spatial and frequency domain channel extrapolation will further reduce the pilot training overhead. However, although some works have pointed out that jointly learning spatial-frequency domain features is beneficial to channel estimation\cite{zhou2024pay,dong2019deep}, spatial-frequency channel extrapolation has not yet been well investigated to reduce the pilot training overhead for mmWave massive MIMO-OFDM systems. In addition, spatial-frequency correlations and long-term temporal domain dependencies (i.e., dependencies over multiple slots) of downlink channels are rarely investigated to reduce computational complexity and facilitate slot-level channel extrapolation. Therefore, we first propose a knowledge-and-data driven spatial-frequency channel extrapolation network (KDD-SFCEN) to reduce the spatial-frequency domain pilot overhead. Specifically, coarse low-dimensional channel estimates are first obtained with a knowledge-driven channel estimator from low-density pilots in spatial and frequency domains and then extrapolated to the full-dimensional CSI with the spatial-frequency channel extrapolation network. After that, a temporal uplink-downlink channel extrapolation network (TUDCEN) is proposed for slot-level channel extrapolation, aiming to reduce the temporal domain pilot overhead. To the best of our knowledge, none of the existing works have investigated such a systematic framework to reduce the pilot training overhead and improve the spectral efficiency of mmWave massive MIMO-OFDM systems under high-mobility scenarios. The major contributions of this paper are summarized as follows:

\begin{enumerate}
\item We propose the KDD-SFCEN to reduce the spatial-frequency domain pilot overhead effectively via joint spatial-frequency channel extrapolation. In addition, it is worth emphasizing that the substantial time and power consumption due to antenna switching can be avoided through spatial extrapolation. The proposed KDD-SFCEN consists of a knowledge-driven coarse channel estimator to provide coarse channel estimates and accelerate the training of the proposed network. The KDD-SFCEN also comprises a spatial-frequency channel extrapolator encompassing the attention-based sub-element extrapolation module (ASEEM) and the progressive extrapolation architecture to exploit spatial and frequency correlations of wireless channels for spatial and frequency domain extrapolation. Specifically, based on the observation that antennas/subcarriers exhibit stronger correlations when they are closer, the progressive extrapolation architecture is designed to first focus on capturing the correlations between neighboring antennas/subcarriers, and then progressively broaden the scope to capture correlations, while the ASEEM is designed to capture specific spatial/frequency correlations in the process.

\item We propose the TUDCEN for accurate slot-level channel extrapolation given the estimated uplink CSI at the first slot. Through the proposed TUDCEN-aided pilot training scheme, channel estimation can be conducted less frequently and more slots can be configured for data transmission than a general pilot training scheme under high-mobility scenarios, further reducing the temporal domain pilot overhead and improving the system's spectral efficiency. The TUDCEN is composed of an uplink-downlink channel calibration network (UDCCN) and a downlink channel extrapolation network (DCEN). The UDCCN calibrates the estimated uplink channel to the downlink channel at the first downlink slot to compensate for the hardware asymmetry in transceivers\cite{jiang2018channel}. The DCEN achieves slot-level channel extrapolation via spatial-frequency sampling embedding and autoregressive generation with the generative Transformer\cite{brown2020language}. Spatial-frequency sampling embedding is proposed to reduce the tremendous computational complexity due to high-dimensional CSI matrices by exploiting spatial and frequency correlations of wireless channels. We exploit the generative Transformer for generating downlink channels autoregressive by learning temporal dependencies of wireless channels over time, which enables slot-level downlink channel extrapolation given only the first downlink channel, thereby avoiding heavy slot-level historical downlink channel estimation.

\item  We use the sounding reference signal (SRS) defined by the 3rd generation partnership project (3GPP) 5G technical specification \cite{3gpp2023ts38211} as the pilot signal for uplink pilot training. The frame structures and system settings in framework design and numerical simulations also follow the 3GPP 5G technical specification. Numerical results demonstrate the superiority of the proposed framework in significantly reducing the spatial-frequency domain pilot overhead by $4$ times via spatial-frequency channel extrapolation compared with state-of-the-art spatial-frequency channel estimation/extrapolation methods. In addition, by reducing the times of single-slot channel estimations with slot-level channel extrapolation, the proposed framework further reduces the temporal domain pilot overhead by $4$ times and significantly improves the mmWave massive MIMO system's spectral efficiency under high-mobility scenarios.

\end{enumerate}

The remainder of this paper is organized as follows. In Section \ref{Sec. Sys.}, we introduce the system model and formulate the spatial, frequency, and temporal channel extrapolation problem. In Section \ref{Sec. UCE}, we propose the KDD-SFCEN for uplink channel estimation to reduce the spatial-frequency domain pilot training overhead. In Section \ref{Sec. TCE}, we propose the TUDCEN for slot-level channel extrapolation to reduce the temporal domain pilot overhead. In Section \ref{Sec. Sim.}, simulation results are presented to demonstrate the superiority of the proposed spatial, frequency, and temporal channel extrapolation framework. Finally, we conclude this work in Section \ref{Sec. Con.}.

\textit{Notation}: Underlined bold uppercase letter $\underline{\mathbf{A}}$, bold uppercase letter $\mathbf{A}$, and bold lowercase letter $\mathbf{a}$ represent a tensor, a matrix, and a vector, respectively. Calligraphy uppercase letter $\mathcal{A}$ represents a set. $\mathbf{A}\{s,t\}$ denotes the representation of $\mathbf{A}$ at the $t$-th slot of the $s$-th sub-frame. $\mathbf{A}_{:,n}$ and $\mathbf{A}_{m,n}$ denote the $n$-th column and the element at the $m$-th row and $n$-th column of the matrix $\mathbf{A}$, respectively. $(\cdot)^{T}$, $(\cdot)^{H}$, and $(\cdot)^{-1}$ denote the transpose, conjugate-transpose, and inverse of a matrix, respectively. $\lceil \cdot \rceil$, $\odot$, $\mathbb{E}\{\cdot\}$, and $\|\cdot\|_2$ denote the ceiling function, Hadamard product, expectation, and L2 norm, respectively.

\begin{figure*}[htbp]
\centering
\includegraphics[width=\textwidth]{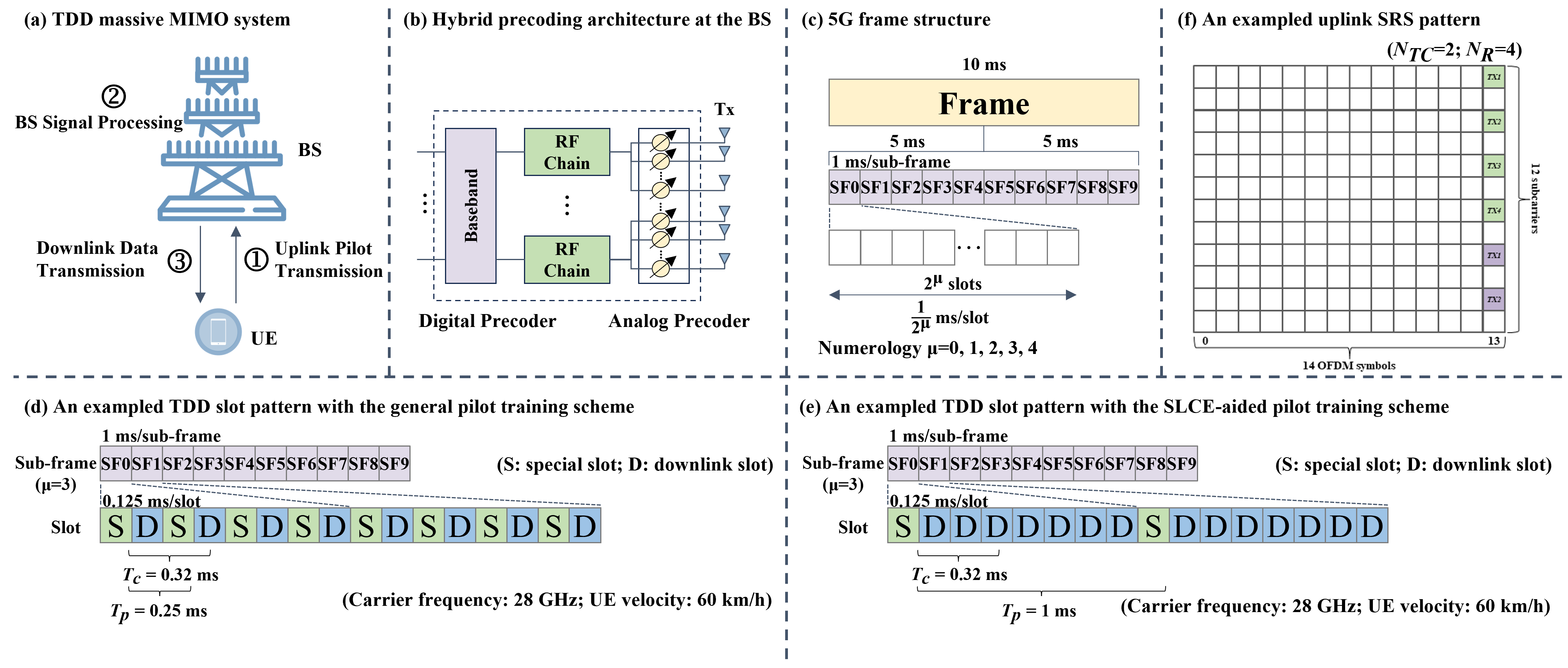}
\caption{A TDD massive MIMO system with the hybrid precoding architecture at the BS. (a) The TDD massive MIMO system; (b) The hybrid precoding architecture at the BS; (c) The 5G frame structure; (d) An exampled TDD slot pattern with a general pilot training scheme; (e) An exampled TDD slot pattern with the SLCE-aided pilot training scheme; (f) An exampled uplink SRS pattern.}
\label{system}
\end{figure*}

\section{System Model and Problem Formulation}\label{Sec. Sys.}

\subsection{System Model}

As shown in Fig. \ref{system} (a), we consider a mmWave massive MIMO system working in the TDD mode, where a single BS equipped with $N_T \gg 1$ antennas and $N_{RF} \ll N_T$ RF chains serves a single user equipped with $N_R$ antennas. The BS is realized with a hybrid precoding architecture, as Fig. \ref{system} (b) shows. The system operates with the OFDM modulation with a total of $N_c \gg 1$ subcarriers. By leveraging the uplink-downlink channel reciprocity in the TDD system, the downlink CSI can be acquired via uplink channel estimation and downlink channel derivation. A simplified downlink CSI acquisition and data transmission process based on uplink pilot training is also shown in Fig. \ref{system} (a). The process consists of three phases, i.e., uplink pilot transmission (phase $1$), BS signal processing (phase $2$), and downlink data transmission (phase $3$). Specifically, the uplink pilot signal is first transmitted by the UE to the BS for uplink channel estimation (phase $1$). Then the BS estimates the uplink channel given the received pilot signal, derives the downlink channel, and conducts downlink precoding (phase $2$). After that, the BS transmits data symbols to the UE based on the downlink precoder and other transmission parameters (phase $3$).\footnote{Note that the working modes of the hybrid array in phase $1$ and phase $3$ are different. In phase $1$, each RF chain is connected to an antenna at a time to receive the individual signal from this antenna and switches to connect to other antennas, which enables CSI acquisition for all antennas and facilitates downlink precoding. While in phase $3$, each RF chain is connected to multiple antennas at the same time and combines the signals received by multiple antennas, achieving a balance between system performance and complexity.}

Due to the limited number of RF chains, only the uplink CSI at $N_{RF}$ antennas connected to the $N_{RF}$ RF chains can be obtained in each uplink pilot signaling process. Therefore, to obtain the uplink CSI at all $N_T$ antennas, the BS has to switch the antennas connected to these $N_{RF}$ RF chains for $\frac{N_T}{N_{RF}}$ times, leading to huge time and power consumption. Specifically, for the $i$-th subcarrier and the $k$-th uplink pilot signaling process, denote the uplink CSI at $N_{RF}$ antennas, the transmitted diagonal pilot signal, the received pilot signal, and the uplink noise as $\mathbf{H}_i^{(k)} \in \mathbb{C}^{N_{RF} \times N_R}$, ${\mathbf{S}^p_i}^{(k)} \in \mathbb{C}^{N_R \times N_R}$, ${\mathbf{Y}^p_i}^{(k)} \in \mathbb{C}^{N_{RF} \times N_R}$, and ${\mathbf{N}^p_i}^{(k)} \in \mathbb{C}^{N_{RF} \times N_R}$, respectively. The set of antennas connected to RF chains in the $k$-th uplink pilot signaling process is denoted as $\mathcal{A}_{RF}^{(k)}$, and $\mathbf{H}_i^{(k)}$ corresponds to the uplink CSI at antennas in $\mathcal{A}_{RF}^{(k)}$. Then, ${\mathbf{Y}^p_i}^{(k)}$ can be represented as
\begin{equation}\label{uplink signal}
{\mathbf{Y}^p_i}^{(k)}  = \mathbf{H}_i^{(k)} {\mathbf{S}^p_i}^{(k)} + {\mathbf{N}^p_i}^{(k)},
\end{equation}
where $\mathbf{H}_i^{(k)}$ can then be estimated by a channel estimator $f_{uce}$ given ${\mathbf{Y}^p_i}^{(k)}$ and ${\mathbf{S}^p_i}^{(k)}$ as
\begin{equation}\label{uplink estimation}
    \hat{\mathbf{H}}_i^{(k)} = f_{uce}({\mathbf{Y}^p_i}^{(k)}, {\mathbf{S}^p_i}^{(k)}),
\end{equation}
where $\hat{\mathbf{H}}_i^{(k)} \in \mathbb{C}^{N_{RF} \times N_R}$ is the estimate of $\mathbf{H}_i^{(k)}$. By repeating the uplink pilot signaling process for $\frac{N_T}{N_{RF}}$ times, the uplink channel at the $i$-th subcarrier, i.e., $\mathbf{H}_i \in \mathbb{C}^{N_T \times N_R}$, can be obtained as
\begin{equation}
\hat{\mathbf{H}}_i = [({\hat{\mathbf{H}}_i^{(1)}})^T, \ldots, ({\hat{\mathbf{H}}_i^{(k)}})^T, \ldots, ({\hat{\mathbf{H}}_i^{(\frac{N_T}{N_{RF}})}})^T]^T,
\end{equation}
where $\hat{\mathbf{H}}_i$ is the estimate of $\mathbf{H}_i$, and
\begin{equation}
\mathcal{A}  = \mathcal{A}_{RF}^{(1)} \cup \ldots \cup \mathcal{A}_{RF}^{(k)} \cup \ldots \cup \mathcal{A}_{RF}^{(\frac{N_T}{N_{RF}})},
\end{equation}
where $\mathcal{A}$ is the set containing all BS antennas. Then, due to the hardware asymmetry in transceivers\cite{jiang2018channel}, the downlink channel at the $i$-th subcarrier, i.e., $\mathbf{H}^d_i \in \mathbb{C}^{N_R \times N_{T}}$, has to be derived from $\hat{\mathbf{H}}_i$ based on the uplink-downlink channel reciprocity and via channel calibration as
\begin{equation}
    \hat{\mathbf{H}}^d_i = f_c(\hat{\mathbf{H}}_i),
\end{equation}
where $\hat{\mathbf{H}}^d_i \in \mathbb{C}^{N_R \times N_T}$ is the estimate of $\mathbf{H}^d_i$, and $f_c(\cdot)$ denotes the uplink-downlink channel calibration function.

After that, hybrid precoding is conducted for downlink transmission and thus the received signal $\mathbf{y}_i \in \mathbb{C}^{N_R \times 1}$ at the $i$-th subcarrier can be represented as
\begin{equation}\label{downlink signal}
    \mathbf{y}_i = \mathbf{H}^d_i \mathbf{F}^{RF}_i \mathbf{F}^{BB}_i \mathbf{s}_i + \mathbf{n}_i,
\end{equation}
where $\mathbf{s}_i \in \mathbb{C}^{N_R \times 1}$ and $\mathbf{n}_i \in \mathbb{C}^{N_R \times 1}$ are the transmitted signal and the downlink noise at the $i$-th subcarrier, and $\mathbf{F}^{RF}_i \in \mathbb{C}^{N_T \times N_{RF}}$ and $\mathbf{F}^{BB}_i \in \mathbb{C}^{N_{RF} \times N_R}$ are the analog and digital precoding matrices, respectively. In addition, $\mathbf{F}^{RF}_i$ can be designed via the discrete Fourier transform (DFT) codebook-based analog precoding
algorithm, and $\mathbf{F}^{BB}_i$ can be designed via the zero-forcing (ZF) algorithm.

\begin{figure*}[htbp]
\centering
\includegraphics[width=\textwidth]{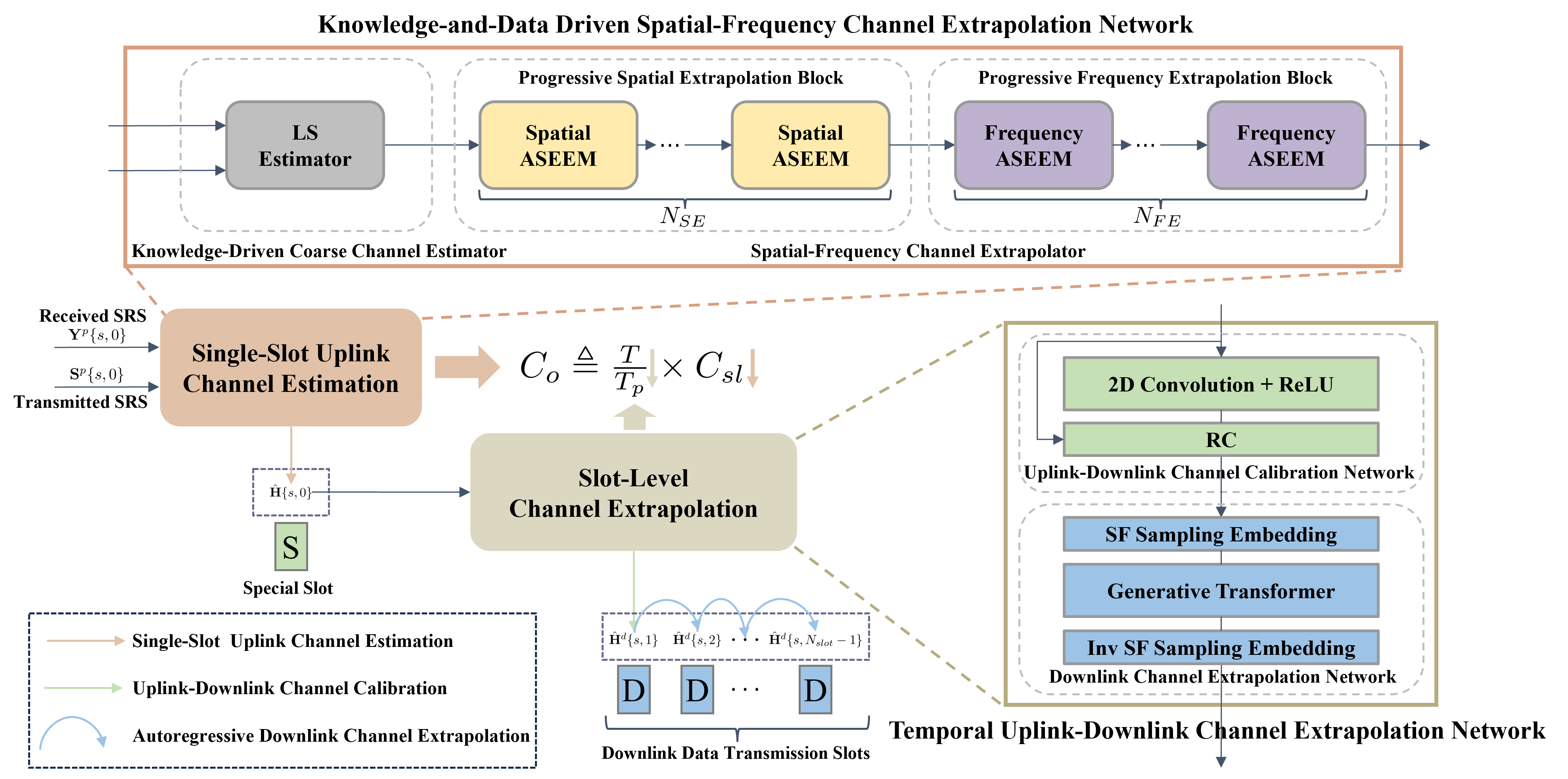}
\caption{The overall framework of the proposed spatial, frequency, and temporal channel extrapolation method. The proposed method first realizes uplink channel estimation with the KDD-SFCEN to reduce the spatial-frequency domain pilot training overhead $C_{sl}$, and then conducts accurate slot-level channel extrapolation with the TUDCEN to reduce the times of uplink channel estimations $\frac{T}{T_p}$, thereby systematically reducing the pilot training overhead $C_o$.}
\label{framework}
\end{figure*}

\subsection{Problem Formulation}
To support the measurement of all transmit ports and subcarriers, the length of the pilot signal is $N_R  \times N_c$. Therefore, the pilot training overhead is $(N_{RF} \times \frac{N_T}{N_{RF}}) \times N_R  \times N_c = N_T \times N_R  \times N_c$, which is extremely high and significantly challenging to the system's spectral efficiency. In addition, the time and power consumption due to the switching of antennas connected to RF chains is also unaffordable. Moreover, the pilot signal should be transmitted for uplink channel estimation every channel coherence time $T_c$ to avoid significant channel aging. For simplicity, we call this scheme the general pilot training scheme hereafter. We denote the pilot signal period as $T_{p}$ and for the general pilot training scheme we have $T_{p} \leq T_c$. Note that as specified in the 3GPP 5G technical specification \cite{3gpp2023ts38211}, the frame structure in the 5G new radio (NR) is shown in Fig. \ref{system} (c). Each frame spans $10$ ms and is divided into two half-frames. Each half-frame spans $5$ ms and contains five sub-frames, with each sub-frame spanning $1$ ms. The number of slots in one sub-frame, i.e., $N_{slot}$, is determined by the numerology $\mu$, and $N_{slot} = 2^\mu$. Here we assume $\mu=3$ which indicates $N_{slot} = 8$ in one sub-frame and each slot spans $0.125$ ms. It should be emphasized that under scenarios with high carrier frequency and high UE mobility, the channel coherence time could be very short. For instance, when operating at a carrier frequency of $28$ GHz and with the UE velocity of $60$ km/h, the channel coherence time is approximately $0.32$ ms\cite{jiang2022accurate}. With the general pilot training scheme, the pilot signal has to be transmitted every two slots (i.e., $T_{p}=0.25$ ms) to mitigate channel aging with $T_c=0.32$ ms as Fig. \ref{system} (d) shows,\footnote{Note that for ease of elaboration, we do not consider uplink data transmission slots in the sub-frame. Our method can be easily extended to sub-frames with uplink data transmission slots. Additionally, in Fig. \ref{system} (d), (e), and (f), we assume the pilot symbol is configured at the last symbol index of each special slot.} further degrading the system's spectral efficiency.

Denote the single-slot pilot training overhead $C_{sl}$ as the pilot training overhead for single-slot uplink channel estimation (i.e., the spatial-frequency domain pilot overhead), which can be mathematically expressed as
\begin{equation}
    C_{sl} = N_{T} \times N_R \times N_c.
\end{equation}

Then, for an interval of length $T$, the uplink channel estimation needs to be conducted for $\frac{T}{T_p}$ times, leading to huge temporal domain pilot overhead. Therefore, the overall pilot training overhead $C_o$ can be defined as
\begin{equation}
    C_o \triangleq \frac{T}{T_p} \times C_{sl}.
\end{equation}

To circumvent such huge overall pilot training overhead, we consider spatial, frequency, and temporal channel extrapolation to systematically reduce the pilot training overhead from two perspectives: 1) reducing the single-slot pilot training overhead $C_{sl}$ via spatial-frequency channel extrapolation; 2) reducing the times of single-slot channel estimations $\frac{T}{T_p}$ to reduce the temporal domain pilot overhead via temporal channel extrapolation. In addition, it is worth noting that through spatial extrapolation, the substantial time and power consumption due to antenna switching can be avoided. And through slot-level channel extrapolation, more slots can be configured for data transmission to improve the system's spectral efficiency. The overall framework of the proposed method is shown in Fig. \ref{framework}.

On the one hand, considering the marginal effects of pilots in the spatial and frequency domains, spatial channel extrapolation and frequency channel extrapolation are combined to reduce the single-slot pilot training overhead $C_{sl}$ and circumvent the time and power consumption during single-slot uplink channel estimation (SSUCE). Specifically, the $N_{RF}$ RF chains are connected to $N_{RF}$ uniformly sampled antennas out of all $N_T$ antennas without antenna switching in the whole uplink channel estimation phase, and the pilot symbols are configured only on partial subcarriers. Ultimately, only the partial uplink CSI can be obtained directly from the received pilot signal, and then the full uplink CSI is extrapolated from the partial uplink CSI. Denote the transmitted diagonal pilot signal matrix and the corresponding received pilot signal matrix by $\mathbf{S}^p \in \mathbb{C}^{(N_R \times N_c^\prime) \times (N_R \times N_c^\prime)}$ and $\mathbf{Y}^p \in \mathbb{C}^{N_{RF} \times (N_R \times N_c^\prime)}$, respectively, where $N_c^\prime$ is the number of subcarriers configured with pilot symbols. In addition, by concatenating $\mathbf{H}_i$, $i=1, 2, \ldots, N_c$, the uplink channel at all subcarriers, denoted by $\mathbf{H} \in \mathbb{C}^{N_T \times (N_R \times N_c)}$, can be given by:
\begin{equation}
    \mathbf{H} = [\mathbf{H}_1, \mathbf{H}_2, \ldots, \mathbf{H}_{N_c}].
\end{equation}
Further denoting $\mathbf{H}\{s,t\} \in \mathbb{C}^{N_T \times (N_R \times N_c)}$, $\mathbf{S}^p\{s,t\} \in \mathbb{C}^{(N_R \times N_c^\prime) \times (N_R \times N_c^\prime)}$, and $\mathbf{Y}^p\{s,t\} \in \mathbb{C}^{N_{RF} \times (N_R \times N_c^\prime)}$ as the uplink channel, the transmitted diagonal pilot signal matrix, and the received pilot signal matrix at the $t$-th slot of the $s$-th sub-frame, respectively, the uplink channel at the first slot of the $s$-th sub-frame can be estimated based on spatial-frequency channel extrapolation as
\begin{equation} \label{P1}
\hat{\mathbf{H}}{\{s,0\}} = F_{ssuce}(\mathbf{Y}^p{\{s,0\}}, \mathbf{S}^p{\{s,0\}}),
\end{equation}
where $\hat{\mathbf{H}}{\{s,0\}}$ is the estimate of $\mathbf{H}{\{s,0\}}$, and $F_{ssuce}$ is a DNN-based SSUCE model.

On the other hand, we propose to reduce the times of single-slot channel estimations $\frac{T}{T_p}$ by leveraging the uplink-downlink channel reciprocity and temporal dependencies for slot-level channel extrapolation (SLCE). In TDD systems, it is commonly assumed that the uplink and downlink channels are reciprocal during the channel coherence time, such that the downlink channel can be derived from the estimated uplink channel. Note that due to high UE mobility and short channel coherence time, the uplink-downlink channel reciprocity only holds for the special slot of a sub-frame and a few downlink data transmission slots in the same sub-frame. And due to the hardware asymmetry in transceivers, uplink-downlink channel calibration is required to derive the downlink channel accurately from the estimated uplink channel. Moreover, like many real-world phenomena, wireless channels exhibit temporal dependencies over time (e.g., explicit patterns like trends, etc). Capturing these temporal dependencies from downlink channels allows us to conduct accurate temporal extrapolation for future slots. However, the temporal dependencies for high-dimensional channels are challenging to capture. Fortunately, with DNNs, it is promising to learn the uplink-downlink channel reciprocity and temporal dependencies for SLCE. Then, with SLCE, the pilot signal can be transmitted with a pilot signal period $T_p \gg T_c$. Taking transmitting the pilot signal only once per sub-frame and at the first slot of each sub-frame as an example, the TDD slot pattern with the SLCE-aided pilot training scheme is provided in Fig. \ref{system} (e).\footnote{Note that for ease of elaboration, in the following context, we simply use this TDD slot pattern and suppose the pilot signal only once per sub-frame. The pilot signal can be set flexibly in practice.} In this example, the pilot signal period is $T_{p} = 1$ ms, which is much larger than the channel coherence time $T_c = 0.32$ ms. To tackle the channel aging issue caused by high carrier frequency and high UE mobility, SLCE is thus anticipated to predict the downlink channels within the pilot signal period accurately for downlink data transmission. Specifically, based on the estimated uplink channel at the special slot, the downlink channel at the first downlink data transmission slot can be initially acquired by exploiting the uplink-downlink channel reciprocity and via channel calibration. Then, by learning the temporal dependencies among time-varying downlink channels, accurate estimates of the downlink channels at all other downlink data transmission slots can be achieved. Therefore, denoting $\mathbf{H}^d\{s,t\} \in \mathbb{C}^{N_R \times (N_T \times N_c)}$ as the downlink channel at the $t$-th slot of the $s$-th sub-frame, the SLCE problem can be formulated as
\begin{equation} \label{P2}
\hat{\mathbf{H}}^d\{s,1\}, \ldots, \hat{\mathbf{H}}^d\{s,N_{slot}-1\}  = F_{slce}(\hat{\mathbf{H}}{\{s,0\}}),
\end{equation}
where $\hat{\mathbf{H}}^d\{s,t\}$ is the estimate of $\mathbf{H}^d\{s,t\}$, $F_{slce}$ is a DNN-based SLCE model.

\subsection{Sounding Reference Signal Pattern}\label{Sec. Sys. C}
Following the 3GPP technical specification\cite{3gpp2023ts38211}, the sounding reference signal (SRS) is used as the pilot signal for uplink pilot training. The SRS is configured according to the transmission comb number $N_{TC}$ in the frequency domain and with frequency domain code division multiplexing (fd-CDM) for multiple antenna ports. For example, assuming the transmission comb number $N_{TC} = 2$ and the UE antenna number $N_R = 4$, the uplink SRS pattern in one resource block (RB) can be illustrated by Fig. \ref{system} (f), where the SRS is configured with every other resource element (RE) and has an fd-CDM length of $4$ to support $4$ UE antennas. Based on $N_{TC}$ and the number of RBs, i.e., $N_{RB}$, the received pilot signal matrix $\mathbf{Y}^p$ is with the size of $N_{RF} \times (N_R \times N_c^\prime) = N_{RF} \times (N_R \times N_{RB} \times \frac{12}{N_{TC}})$, where $12$ is the number of consecutive subcarriers that form a RB as defined in \cite{3gpp2023ts38211}. Therefore, compared with the size of the uplink channel $\mathbf{H}$, i.e., $N_T \times (N_R \times N_c) = N_T \times (N_R \times N_{RB} \times 12)$, we define the spatial compression ratio as $R_s = \frac{N_T}{N_{RF}}$ and the frequency compression ratio as $R_f = N_{TC}$.

In the following two sections, to reduce the overall pilot training overhead $C_o$ from the aforementioned two perspectives, i.e., reducing the single-slot pilot training overhead $C_{sl}$ and reducing the times of single-slot channel estimations $\frac{T}{T_p}$, we propose the KDD-SFCEN for single-slot uplink channel estimation and the TUDCEN for slot-level channel extrapolation respectively.

\section{Spatial-Frequency Channel Extrapolation based Single-Slot Uplink Channel Estimation}\label{Sec. UCE}
Considering the marginal effects of pilots in the spatial and frequency domains, we propose to conduct joint spatial-frequency channel extrapolation to reduce the pilot training overhead of single-slot uplink channel estimation. Traditional channel estimation methods based on domain knowledge have been fully validated for decades that they can provide feasible channel estimates in time-invariant channels, which can be exploited as initial estimates for deep learning-based channel estimation methods. In addition, to estimate the uplink channel accurately with only a few pilots in both the spatial and frequency domains, the spatial and frequency characteristics of massive MIMO channels must be fully exploited \cite{zhou2024pay}. Therefore, in this section, we propose the KDD-SFCEN for single-slot uplink channel estimation in TDD massive MIMO systems, which benefits from both domain knowledge via the traditional knowledge-driven channel estimation and data via joint spatial-frequency channel extrapolation. As shown in Fig. \ref{framework}, the proposed network consists of a knowledge-driven coarse channel estimator and a spatial-frequency channel extrapolator for coarse channel estimation and spatial-frequency channel extrapolation, respectively.

\subsection{Coarse Channel Estimation}
Since traditional LS estimation has been demonstrated to be a simple but effective method to achieve coarse channel estimates, we exploit the traditional LS estimator as the knowledge-driven coarse channel estimator to accelerate the training of our proposed uplink channel estimation network. Specifically, the received pilot signal $\mathbf{Y}^p{\{s,0\}}$ and the corresponding transmitted diagonal pilot signal $\mathbf{S}^p{\{s,0\}}$ is first processed by an LS estimator as
\begin{equation}
\hat{\mathbf{H}}^{LS} = \mathbf{Y}^p{\{s,0\}}\{\mathbf{S}^p{\{s,0\}}\}^{-1},
\end{equation}
where $\hat{\mathbf{H}}^{LS} \in \mathbb{C}^{N_{RF} \times (N_R \times N_c^\prime)}$ is the coarse channel estimate achieved by the knowledge-driven coarse channel estimator.

Then, the coarse channel estimate is further extrapolated and refined by the spatial and frequency channel extrapolator to achieve a highly accurate channel estimate.

\begin{figure}[htbp]
\centering
\includegraphics[width=0.8\columnwidth]{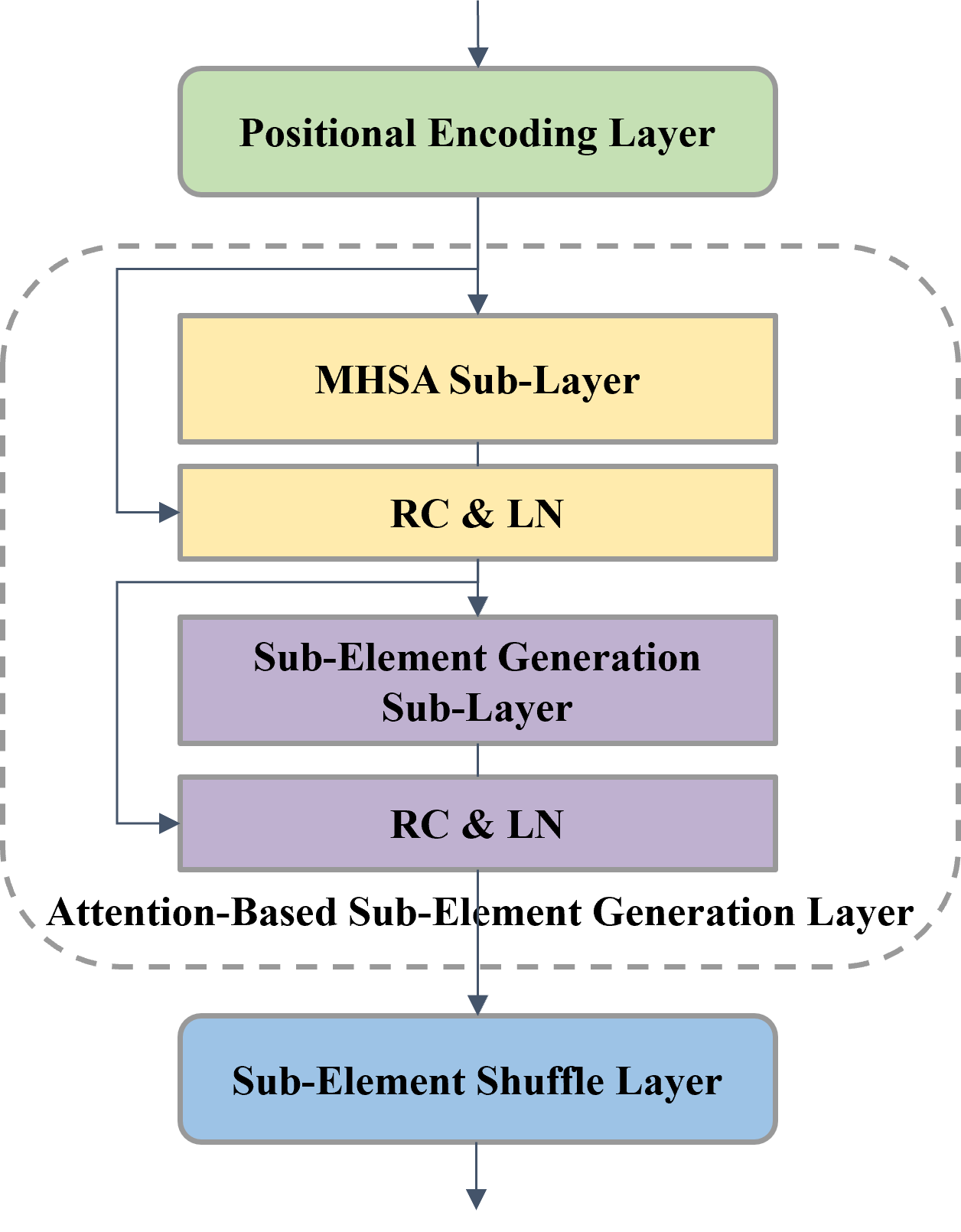}
\caption{The architecture of the proposed ASEEM.}
\label{aseem}
\end{figure}

\subsection{Spatial and Frequency Channel Extrapolation}
To estimate the uplink channel accurately with a few pilots, we propose the spatial-frequency channel extrapolator consisting of ASEEMs and the progressive extrapolation architecture to exploit spatial and frequency correlations of wireless channels progressively for spatial and frequency domain extrapolation. The spatial correlation refers to the correlation between the signals received at different antennas in the antenna domain, which practically depends on the spatial characteristics of the environment (e.g., multipath components, angle of arrival, etc.) and antenna configurations (e.g., antenna geometry, antenna spacing, etc). The frequency correlation refers to the correlation of the channel across different subcarriers in the frequency domain, which is mainly influenced by the channel’s multipath delay spread and Doppler shift due to mobility. In addition, it can also be observed that the closer the antennas/subcarriers are, the stronger the correlations among them are. Based on these observations, we design the progressive extrapolation architecture to first focus on capturing the correlations between neighboring antennas/subcarriers, and then progressively broaden the scope to capture correlations, with the ASEEM designed to capture specific spatial/frequency correlations in the process. As Fig. \ref{framework} shows, the proposed spatial-frequency channel extrapolator comprises a progressive spatial extrapolation block with $N_{SE}$ spatial ASEEMs and a progressive frequency extrapolation block with $N_{FE}$ frequency ASEEMs for spatial and frequency extrapolation, respectively. These spatial/frequency ASEEMs progressively extrapolate partial spatial/frequency CSI to full spatial/frequency CSI.

\subsubsection{ASEEMs} Partial CSI in either the spatial or frequency domain can be regarded as downsampled information from full CSI in the specific domain, which is similar to an image taken from the real world. Generally, a pixel is the smallest addressable element of an image. However, a pixel in an image might correspond to many small objects in the real world. Motivated by this, sub-pixel imaging technologies emerge by dividing a single pixel into smaller sub-pixels to improve the resolution of an image. Similarly, learning to divide each element in a partial CSI into smaller sub-elements will facilitate CSI extrapolation. This motivates us to propose the ASEEM, which is composed of one positional encoding layer, one attention-based sub-element generation layer, and one sub-element shuffle layer. To better illustrate the proposed ASSEM, the architecture of the ASEEM is provided in Fig. \ref{aseem}.

Denote the input to the ASEEM as $\mathbf{X} \in \mathbb{R}^{N_I \times d_R}$, where $N_I$ is the number of initial input elements and $d_R$ is the representation dimension of each input element. The positional encoding layer generates a positional encoding (PE) $\mathbf{P} \in \mathbb{R}^{N_I \times d_R}$ for all input elements of $\mathbf{X}$ as
\begin{align}
\mathbf{P} &= \operatorname{Embed}(\mathbf{p}^{idx};\Theta^P), \label{PE1}\\
\mathbf{p}^{idx} &= [1,2,\ldots,N_I], \label{PE2}
\end{align}
where $\mathbf{p}^{idx} \in \mathbb{R}^{N_I \times 1}$ is the positional index vector to all input elements, $\operatorname{Embed}(\cdot)$ is the embedding function parameterized by $\Theta^P \in \mathbb{R}^{N_I \times d_R}$ that converts the positional index vector $\mathbf{p}^{idx}$ into the learnable positional encoding $\mathbf{P}$. Then, the input to the attention-based sub-element generation layer, denoted by $\tilde{\mathbf{X}}  \in \mathbb{R}^{N_I \times d_R}$ can be obtained by
\begin{equation}
    \tilde{\mathbf{X}} = \mathbf{X} + \mathbf{P}.
\end{equation}

The attention-based sub-element generation layer further consists of one multi-head self-attention (MHSA) sub-layer, one sub-element generation (SEG) sub-layer, and residual connection (RC)\cite{he2016deep} and layer normalization (LN)\cite{ba2016layer} operations around each of the two sub-layers. By splitting the representation of the input elements into several representation sub-spaces, the MHSA mechanism allows the neural network to capture diverse aspects of the input elements simultaneously, leading to better representation learning compared to the single-head self-attention (SHSA) mechanism. The SEG sub-layer ensures the non-linear representation capability of the attention-based sub-element generation layer and generates sub-elements for each input element, and the RC and LN operations stabilize the neural network training. Specifically, $\tilde{\mathbf{X}}$ is first projected to query, key, and value matrices, i.e., $\mathbf{Q} \in \mathbb{R}^{N_I \times d_R}$, $\mathbf{K} \in \mathbb{R}^{N_I \times d_R}$, and $\mathbf{V} \in \mathbb{R}^{N_I \times d_R}$, and then these three matrices are split into $N_h$ sub-matrices, respectively. For the $i$-th ($i=1,2,\ldots,N_h$) sub-matrix set, i.e., $\{\mathbf{Q}_i \in \mathbb{R}^{N_I \times d_Q}, \mathbf{K}_i \in \mathbb{R}^{N_I \times d_K}, \mathbf{V}_i \in \mathbb{R}^{N_I \times d_V}\}$, with $d_Q = d_K = d_V = d_R/N_h$, a self-attention score matrix $\mathbf{Z}_i \in \mathbf{R}^{N_I \times N_I}$ is obtained via the scaled dot-product attention (SDPA) \cite{vaswani2017attention} as
\begin{equation} \label{score}
\mathbf{Z}_i = \operatorname{Drop}\left(\frac{\operatorname{exp}(\frac{\mathbf{Q}_i \mathbf{K}_i^{T}}{\sqrt{d_K}})}{\sum_{j=1}^{N_I} \operatorname{exp}({\frac{\mathbf{Q}_i \mathbf{K}_i^{T}}{\sqrt{d_K}}})_{:,j}};p_1\right),
\end{equation}
where $\operatorname{Drop}(\cdot;p)$ denotes the dropout function with a dropout probability $p$ \cite{srivastava2014dropout}, and the attention-weighted output of the $i$-th head, denoted by $\mathbf{O}_i \in \mathbf{R}^{N_I \times d_V}$, is obtained as
\begin{equation} \label{weighted_output}
\mathbf{O}_i = \mathbf{Z}_i \mathbf{V}_i.
\end{equation}
Finally, the attention-weighted outputs of all $N_h$ heads are concatenated and projected to form the output of the MHSA sub-layer, i.e., $\mathbf{X}^{MHSA} \in \mathbb{R}^{N_I \times d_R}$, as
\begin{align}
\mathbf{X}^{MHSA} &= [\mathbf{O}_{1}, \mathbf{O}_{2}, \ldots, \mathbf{O}_{N_h}]\mathbf{W}^{O},
\end{align}
where $\mathbf{W}^{O} \in \mathbb{R}^{d_R \times d_R}$ is a learnable weight matrix. The outputs of the RC and LN operations around the MHSA sub-layer, denoted by $\mathbf{X}^{RC1} \in \mathbb{R}^{N_I \times d_R}$ and $\mathbf{X}^{LN1} \in \mathbb{R}^{N_I \times d_R}$, respectively, can be formulated by
\begin{equation}
\mathbf{X}^{RC1} = \mathbf{X}^{MHSA} + \tilde{\mathbf{X}}, \label{RC1_1}
\end{equation}
and
\begin{align} \label{ln1}
\mathbf{X}^{LN1} &= \frac{\mathbf{X}^{RC1}-\mu_{\mathbf{X}^{RC1}}}{\sigma_{\mathbf{X}^{RC1}}} \odot \mathbf{g}^{LN1} + \mathbf{b}^{LN1},\\
\mu_{\mathbf{X}^{RC1}} &= \frac{1}{d_R}\sum_{i=1}^{d_R} \mathbf{X}^{RC1}_{:,i},\\
\sigma_{\mathbf{X}^{RC1}} &= \sqrt{\frac{1}{d_R}\sum_{i=1}^{d_R} (\mathbf{X}^{RC1}_{:,i} - \mu_{\mathbf{X}^{RC1}})^2}, \label{LN1_3}
\end{align}
where $\mu_{\mathbf{X}^{RC1}} \in \mathbb{R}^{N_I \times 1}$ and $\sigma_{\mathbf{X}^{RC1}} \in \mathbb{R}^{N_I \times 1}$ denote the mean and standard deviation of $\mathbf{X}^{RC1}$, $\mathbf{g}^{LN1} \in \mathbb{R}^{1 \times d_R}$ and $\mathbf{b}^{LN1} \in \mathbb{R}^{1 \times d_R}$ are learnable affine transformation parameters, and $\odot$ is the Hadamard product.
The output of the SEG sub-layer, i.e., $\mathbf{X}^{G} \in \mathbb{R}^{N_I \times (r \times d_R)}$, can be formulated by
\begin{equation}\label{SEG}
\mathbf{X}^{G} = (\operatorname{Drop}(\operatorname{ReLU}(\mathbf{X}^{RC1}\mathbf{W}^{G1}+\mathbf{b}^{G1});p_2))\mathbf{W}^{G2}+\mathbf{b}^{G2},
\end{equation}
where $p_2$ is a dropout probability, $\operatorname{ReLU}(\cdot)$ is the rectified linear unit (ReLU) activation function, $\mathbf{W}^{G1} \in \mathbb{R}^{d_R \times d_{G}}$ and $\mathbf{W}^{G2} \in \mathbb{R}^{d_{G} \times (r \times d_R)}$ are learnable weight matrices where $d_{G}=d_{R}$, and $\mathbf{b}^{G1} \in \mathbb{R}^{1 \times d_{G}}$ and $\mathbf{b}^{G2} \in \mathbb{R}^{1 \times (r \times d_R)}$ are learnable bias vectors. $r$ is the number of sub-elements to be generated, which is also known as the upscale factor. The output of the RC operation around the SEG sub-layer, i.e., $\mathbf{X}^{RC2} \in \mathbb{R}^{N_I \times (r \times d_R)}$, can be formulated by
\begin{equation}
\mathbf{X}^{RC2} = \mathbf{X}^{G} + \mathbf{X}^{LN1}\mathbf{W}^{RC2}, \label{RC2_1}
\end{equation}
where $\mathbf{W}^{RC2} \in \mathbb{R}^{d_R \times (r \times d_R)}$ is a learnable weight matrix. The output of the LN operation around the SEG sub-layer, i.e., $\mathbf{X}^{LN2} \in \mathbb{R}^{N_I \times (r \times d_R)}$, can be formulated by
\begin{align} \label{ln2}
\mathbf{X}^{LN2} &= \frac{\mathbf{X}^{RC2}-\mu_{\mathbf{X}^{RC2}}}{\sigma_{\mathbf{X}^{RC2}}} \odot \mathbf{g}^{LN2} + \mathbf{b}^{LN2},\\
\mu_{\mathbf{X}^{RC2}} &= \frac{1}{d_R}\sum_{i=1}^{d_R} \mathbf{X}^{RC2}_{:,i},\\
\sigma_{\mathbf{X}^{RC2}} &= \sqrt{\frac{1}{d_R}\sum_{i=1}^{d_R} (\mathbf{X}^{RC2}_{:,i} - \mu_{\mathbf{X}^{RC2}})^2}, \label{LN2_3}
\end{align}
where $\mu_{\mathbf{X}^{RC2}} \in \mathbb{R}^{N_I \times 1}$ and $\sigma_{\mathbf{X}^{RC2}} \in \mathbb{R}^{N_I \times 1}$ denote the mean and standard deviation of $\mathbf{X}^{RC2}$, $\mathbf{g}^{LN2} \in \mathbb{R}^{1 \times (r \times d_R)}$ and $\mathbf{b}^{LN2} \in \mathbb{R}^{1 \times (r \times d_R)}$ are learnable affine transformation parameters.

After the attention-based sub-element generation layer generates $r$ sub-elements for each input element, the sub-element shuffle layer shuffles these sub-elements together to form a new input element matrix as
\begin{equation}
\mathbf{X}^E = [\operatorname{R}(\mathbf{X}^{LN2}_{:,1:r}), \operatorname{R}(\mathbf{X}^{LN2}_{:,r+1:2r}), \ldots, \operatorname{R}(\mathbf{X}^{LN2}_{:,r \times d_R - r + 1:r \times d_R})],
\end{equation}
where $\mathbf{X}^{E} \in \mathbb{R}^{(r \times N_I) \times d_R}$, and $\operatorname{R}$ denotes a rearrange operator that transforms the size of $\mathbf{X}^{LN2}_{:,i:i+r-1} \in \mathbb{R}^{N_I \times r}$ into $(r \times N_I) \times 1$. Therefore, through the ASEEM, the original input $\mathbf{X} \in \mathbb{R}^{N_I \times d_R}$ can be extrapolated to $\mathbf{X}^{E} \in \mathbb{R}^{(r \times N_I) \times d_R}$ effectively via to sub-element generation.

\subsubsection{Progressive Extrapolation Architecture} Since directly extrapolating partial CSI to full CSI is significantly challenging, especially when the spatial and frequency compression ratios are large, we propose to conduct spatial and frequency channel extrapolation progressively with the progressive extrapolation architecture. The spatial extrapolation and frequency extrapolation are organized in a sequential manner, i.e., the $N_{SE}$ spatial ASEEMs of the progressive spatial extrapolation block first progressively extrapolate partial spatial CSI to full spatial CSI, and then the $N_{FE}$ frequency ASEEMs of the progressive frequency extrapolation block progressively extrapolate partial frequency CSI to full frequency CSI. During this process, the sizes of the inputs to the $i$-th spatial ASEEM, i.e., $\mathbf{X}^{S}(i)$, and the $j$-th frequency ASEEM, i.e., $\mathbf{X}^{F}(i)$, progressively increase as
\begin{align}
    \mathbf{X}^{S}(i) \in \mathbb{R}^{(r_s^{i-1} \times N_{SI}) \times d_{SR}},\ & i = 1,2,\ldots,N_{SE}, \\
    \mathbf{X}^{F}(j) \in \mathbb{R}^{(r_f^{j-1} \times N_{FI}) \times d_{FR}},\ & j = 1,2,\ldots,N_{FE},
\end{align}
where $r_s$,  $N_{SI}$, and $d_{SR}$ are the spatial upscale factor, the number of initial spatial input elements, and the spatial representation dimension, respectively; and $r_f$, $N_{FI}$, and $d_{FR}$ are the frequency upscale factor, the number of initial frequency input elements, and the frequency representation dimension, respectively. And $N_{SE}$ and $N_{FE}$ are determined by
\begin{align}
    N_{SE} &= \lceil\log_{r_s}^{R_s} \rceil, \\
    N_{FE} &= \lceil\log_{r_f}^{R_f} \rceil,
\end{align}
where $\lceil \cdot \rceil$ denotes the ceiling function. Correspondingly, the sizes of the outputs of the $i$-th spatial ASEEM, i.e., $\mathbf{X}^{SE}(i)$, and the $j$-th frequency ASEEM, i.e., $\mathbf{X}^{FE}(i)$, progressively increase as
\begin{align}
    \mathbf{X}^{SE}(i) \in \mathbb{R}^{(r_s^{i} \times N_{SI}) \times d_{SR}},\ & i = 1,2,\ldots,N_{SE}, \\
    \mathbf{X}^{FE}(j) \in \mathbb{R}^{(r_f^{j} \times N_{FI}) \times d_{FR}},\ & j = 1,2,\ldots,N_{FE}.
\end{align}
In addition, the sizes of the spatial positional encoding (SPE) for the $i$-th spatial ASEEM and the frequency positional encoding (FPE) for the $j$-th frequency ASEEM progressively increase as
\begin{align}
    \mathbf{P}^{S}(i) \in \mathbb{R}^{(r_s^{i-1} \times N_{SI}) \times d_{SR}},\ & i = 1,2,\ldots,N_{SE}, \\
    \mathbf{P}^{F}(j) \in \mathbb{R}^{(r_f^{j-1} \times N_{FI}) \times d_{FR}},\ & j = 1,2,\ldots,N_{FE}.
\end{align}

\subsection{Model Training}

The proposed KDD-SFCEN is trained via the mean squared error (MSE) loss defined as\footnote{The inputs to $F_{ssuce}$ are actually real-valued tensors transformed from $\mathbf{Y}^p{\{s,0\}}$ and $\mathbf{S}^p{\{s,0\}})$. We use their original complex-valued matrix form in the loss expression for simplicity.}
\begin{equation}\label{L1}
L_1 = \mathbb{E}_{N_{train}} \left(\| \mathbf{H}{\{s,0\}} - F_{ssuce}(\mathbf{Y}^p{\{s,0\}},\mathbf{S}^p{\{s,0\}})\|_2^2\right),
\end{equation}
where $\mathbb{E}_{N_{train}}(\cdot)$ denotes the expectation over $N_{train}$ training samples, and $\|\cdot\|_2$ denotes the L2 norm.

\textit{Remark:} It can be noticed that many works on mmWave channel estimation estimate sparse parameters of mmWave channels exploiting channel sparsity. However, we don't intentionally introduce channel sparsity to design our 3D channel extrapolation framework. The reasons are two-fold. On the one hand, exploiting the sparsity would require some prior knowledge of the channels' specific sparse representation (e.g., the angular spread of angle of arrival (AoA) and angle of departure (AoD)). Although mmWave channels generally exhibit sparsity, in some indoor mmWave communication scenarios with rich multipath components, it is difficult to obtain an accurate sparse representation of the channels, leading to degraded channel estimation performance of channel sparsity-based channel estimation methods. On the other hand, even in non-mmWave systems without channel sparsity, our 3D channel extrapolation framework is expected to work well to reduce the power consumption of RF chains. Therefore, channel sparsity is not incorporated into the design of the 3D channel extrapolation framework to generalize its application scenarios. Nonetheless, it is worth noting that for scenarios with explicit channel sparsity, the 3D channel extrapolation framework can be easily extended to exploit channel sparsity to further reduce the pilot overhead and computational complexity. Specifically, the 3D channel extrapolation framework is currently trained to estimate the full channel matrix given the supervision of the full channel matrix. Suppose prior knowledge of the channel's sparse representation is available, the 3D channel extrapolation framework can be trained to first estimate sparse parameters and then reconstruct the full channel matrix with two-level supervision given labels of both sparse parameters and the full channel matrix.

After obtaining the estimated uplink channel, the slot-level channel extrapolation can then be conducted to reduce the times of uplink channel estimations $\frac{T}{T_p}$.

\section{Channel Reciprocity and Temporal Dependency Aided Slot-Level Channel Extrapolation}\label{Sec. TCE}

An accurate initial state is essential to temporal channel extrapolation. In TDD systems, the downlink channel of the first downlink slot of a sub-frame can be accurately derived from the uplink channel of the sub-frame's special slot via uplink-downlink channel calibration. This provides an accurate initial state for predicting future downlink channels. Recent advances in generative artificial intelligence (GAI), such as ChatGPT\cite{openai2022introducing}, have demonstrated the power of autoregressive models like generative Transformers\cite{brown2020language} in handling sequential data. Given that time-varying channels also form sequential data, generative Transformers are promising on account of their capabilities in learning complicated temporal dependencies and autoregressive modeling. Accordingly, we propose the TUDCEN, comprising the UDCCN and the DCEN, for slot-level channel extrapolation. The architecture of the TUDCEN is shown in Fig. \ref{framework}. Specifically, the UDCCN handles uplink-downlink calibration for initialization, while the DCEN uses the proposed spatial-frequency sampling embedding module to reduce computational complexity and generative Transformers for temporal dependency learning and autoregressive channel extrapolation.

\subsection{Channel Reciprocity Aided Uplink-Downlink Channel Calibration}
Although the physical uplink-downlink channels within the channel coherence time are generally reciprocal, uplink-downlink channel calibration is practically necessary to compensate for the hardware asymmetry in transceivers\cite{jiang2018channel}. Specifically, downlink and uplink channels are affected by the transmit and receive hardware (e.g., amplifiers, filters, mixers), respectively. The hardware impairments, such as amplifier nonlinearities, in-phase (I) / quadrature (Q) imbalances, or phase noise, can make the practical downlink and uplink channel responses non-reciprocal. These hardware impairments affect the ideal reciprocity and thus require uplink-downlink channel calibration to adjust for these mismatches. The uplink-downlink channel extrapolation can be formulated by
\begin{align}
\hat{\mathbf{H}}^d\{s,1\} = F_{udcc}(\hat{\mathbf{H}}{\{s,0\}}),
\end{align}
where $F_{udcc}$ denotes the UDCCN.

To calibrate the downlink channel from the estimated uplink channel, both the spatial and frequency characteristics of the channels should be exploited. Therefore, we propose the two-dimensional (2D) convolution-based calibration network for uplink-downlink channel calibration. The 2D convolution operation can capture spatial and frequency characteristics for accurate calibration. In addition, the ReLU activation function and the RC operation are also applied to improve the network's non-linear representation capability and stabilize the network training, respectively. The input to the UDCCN is obtained by
\begin{align}
\underline{\mathbf{X}}^U &= f_{\mathbb{C} \to \mathbb{R}}(\hat{\mathbf{H}}{\{s,0\}}),
\end{align}
where $\underline{\mathbf{X}}^U \in \mathbb{R}^{(N_T \times N_R) \times N_c \times 2}$ is a real-valued tensor form representation of the estimated uplink channel and the $2$ in the dimensions comes from the concatenated real and imaginary parts of the original complex-valued matrix, and $f_{\mathbb{C} \to \mathbb{R}}$ is a function to transform the original complex-valued matrix form representation to the real-valued tensor form representation. The output of the UDCCN, denoted by $\underline{\mathbf{X}}^D \in \mathbb{R}^{(N_T \times N_R) \times N_c \times 2}$, can be formulated by
\begin{align}
\tilde{\underline{\mathbf{X}}}^D &= \operatorname{ReLU}{(\operatorname{Conv}^{k \times k}(\underline{\mathbf{X}}^U);\mathbf{W}^{Conv})} + \underline{\mathbf{X}}^U, \\
\underline{\mathbf{X}}^D &= \tilde{\underline{\mathbf{X}}}^D \mathbf{W}^D
\end{align}
where $\operatorname{Conv}^{k \times k}$ is a 2D convolution operation with kernel size being $k \times k$ and parameterized by $\mathbf{W}^{Conv} \in \mathbb{R}^{(2k^2) \times d_f}$, $d_f$ is the feature dimension, $\tilde{\underline{\mathbf{X}}}^D \in \mathbb{R}^{(N_T \times N_R) \times N_c \times (2+d_f)}$ is the intermediate feature representation, and $\mathbf{W}^D \in \mathbb{R}^{(2+d_f) \times 2}$ is a learnable weight matrix that projects the intermediate feature representation to the final output $\underline{\mathbf{X}}^D$. Then, the calibrated downlink channel can be obtained by
\begin{align}
\hat{\mathbf{H}}^d{\{s,1\}} &= f_{\mathbb{R} \to \mathbb{C}}(\underline{\mathbf{X}}^D),
\end{align}
where $f_{\mathbb{R} \to \mathbb{C}}$ is a function to transform the real-valued tensor form representation to the original complex-valued matrix form representation.

\subsection{Temporal Dependency Learning and Downlink Channel Extrapolation}
It is worth mentioning that the downlink channel extrapolation problem is conducted along the time domain and aims to predict channels at future slots, which can be regarded as a time series prediction problem. However, different from general time series prediction problems, historical observations in the investigated downlink channel extrapolation problem (i.e., downlink channels at all downlink slots) can not be obtained unless huge time-frequency resources were spent to conduct channel estimation for all downlink slots. As a result, the performance of widely adopted non-autoregressive prediction schemes in related works\cite{jiang2022accurate} would be significantly degraded due to the lack of available historical channel information. Note that like many real-world phenomena, wireless channels exhibit temporal dependencies over time. Then, based on the chain rule in the probability theory, the joint probability of the concatenated downlink channels can be expressed as
\begin{align}\label{chain}
&P([\hat{\mathbf{H}}^d\{s,1\}, \hat{\mathbf{H}}^d\{s,2\}, \ldots, \hat{\mathbf{H}}^d\{s,N_{slot}-1\}]) = \nonumber \\
& P(\hat{\mathbf{H}}^d\{s,1\}) \times P(\hat{\mathbf{H}}^d\{s,2\} | \hat{\mathbf{H}}^d\{s,1\}) \times \ldots \times \nonumber \\
& P(\hat{\mathbf{H}}^d\{s,N_{slot}-1\} | \hat{\mathbf{H}}^d\{s,1\} \hat{\mathbf{H}}^d\{s,2\} \hat{\mathbf{H}}^d\{s,N_{slot}-2\}).
\end{align}
Inspired by this, the objective of the downlink channel extrapolation is
\begin{align}\label{autoregressive}
\hat{\mathbf{H}}^d\{s,t\} &= F_{dce}([\hat{\mathbf{H}}^d\{s,1\},\ldots,\hat{\mathbf{H}}^d\{s,t-1\}]), \nonumber \\
\text{for}\ \ t &= 2, 3, \ldots, N_{slot} - 1,
\end{align}
where $F_{dce}$ denotes the DCEN, such that (\ref{chain}) can be achieved and the downlink channels from the $2$nd slot to the ($N_{slot}-1$)-th slot can be extrapolated given only $\hat{\mathbf{H}}^d\{s,1\}$. It is worth noting that (\ref{autoregressive}) actually refers to the autoregressive generation manner and the DCEN should thus be an autoregressive model. And considering that $\hat{\mathbf{H}}^d\{s,t\}, t=1, 2, \ldots, N_{slot} - 1$ are dependent, capturing these temporal dependencies among them would be beneficial to achieve (\ref{autoregressive}). To this end, we propose the DCEN constituted by (inverse) spatial-frequency sampling embedding layers to reduce computational complexity and generative Transformers for temporal dependency learning and autoregressive downlink channel extrapolation. The architecture of the DCEN is also shown in Fig. \ref{framework}.

\begin{figure*}[htbp]
\centering
\includegraphics[width=\textwidth]{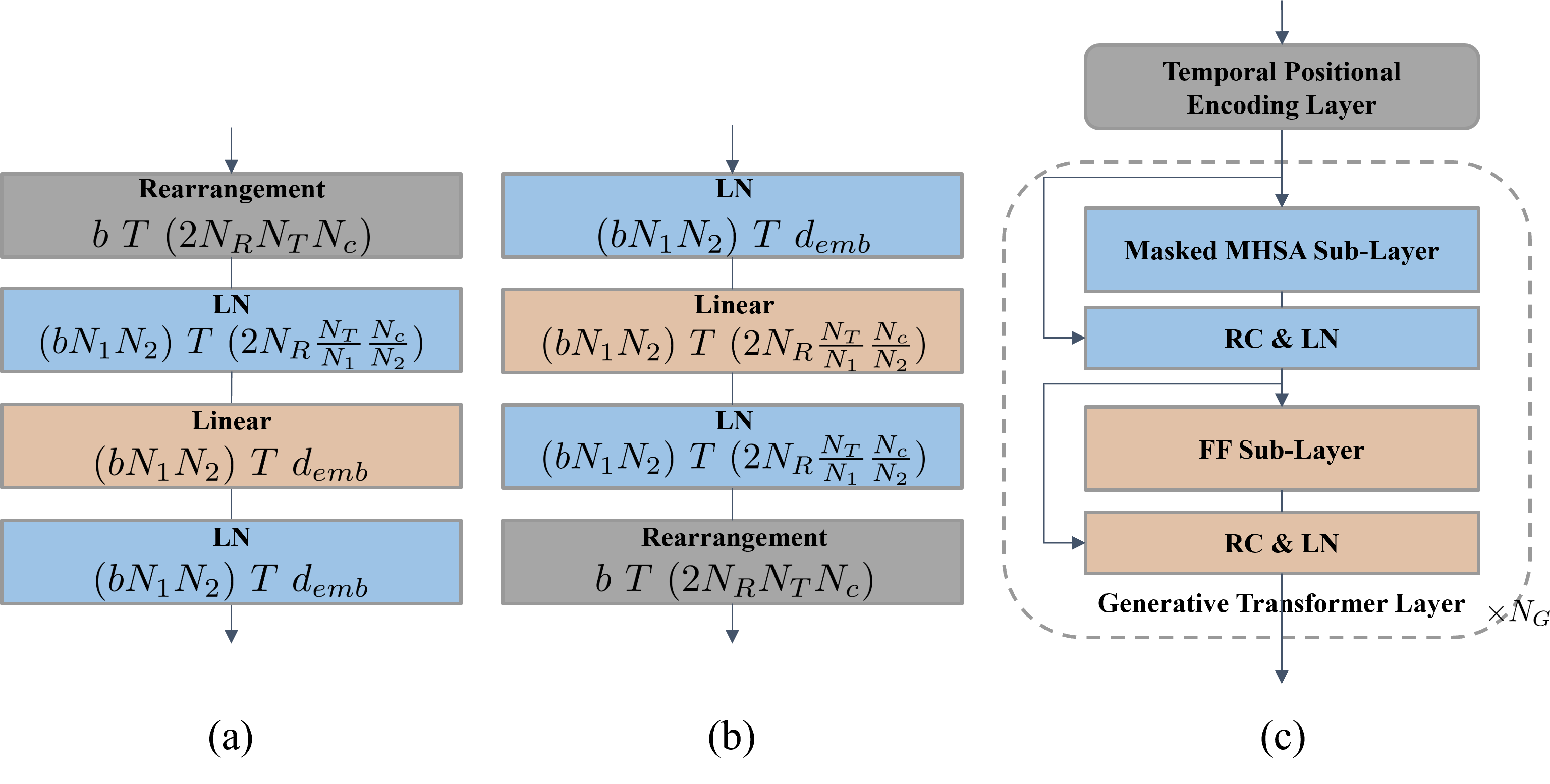}
\caption{The architecture of the (inverse) spatial-frequency sampling embedding layers and the generative Transformer.}
\label{DCE}
\end{figure*}

\subsubsection{Spatial-Frequency Sampling Embedding} Due to the large number of antennas and subcarriers, the downlink channels' spatial and frequency domain representation must be compressed to reduce the computational complexity of downlink channel extrapolation. Therefore, resorting to spatial and frequency correlations of wireless channels, we propose the spatial-frequency sampling embedding layer that can reduce the spatial and frequency dimensions of high-dimensional CSI matrices, thereby reducing the computational complexity. The spatial-frequency sampling embedding layer consists of one rearrangement operation, two LN operations, and one linear layer, as Fig. \ref{DCE} (a) shows. The rearrangement operation splits the $N_T$ transmit antennas and $N_c$ subcarriers into $N_1$ transmit antenna groups and $N_2$ subcarrier groups, respectively. Each transmit antenna group has $\frac{N_T}{N_1}$ antennas and each subcarrier group has $\frac{N_c}{N_2}$ subcarriers. Note that since the number of UE antennas, i.e., $N_R$, is generally small, it is not necessary to further split UE antennas. Specifically, the rearrangement operation rearranges the sample dimension (known as the batch size $N_b$), $N_1$, and $N_2$ together to form a new sample dimension, and consequently reforms a representation dimension of length $2 \times N_R \times \frac{N_T}{N_1} \times \frac{N_c}{N_2}$. Then, the dimension of the reformed representation is normalized by the first LN operation. Following that, the dimension is further projected to an embedding dimension $d_{emb}$ by the linear layer and normalized by the second LN operation. Straightforwardly, the spatial-frequency sampling embedding layer samples the spatial and frequency domain representation of downlink channels with a spatial sampling factor $N_1$ and a frequency sampling factor $N_2$ to reduce the representation dimension by resorting to the spatial and frequency correlations of downlink channels. Meanwhile, the $N_1$ sampled transmit antenna groups and $N_2$ sampled subcarrier groups are combined to facilitate model training. It is worth noting that after the generative Transformers, the inverse spatial-frequency sampling embedding layer is applied to transform the sampled representation dimension back to the original spatial and frequency domain representation. In the inverse spatial-frequency sampling embedding layer, all layers and operations are placed and operated in the inverse direction to the spatial-frequency sampling embedding layer, as Fig. \ref{DCE} (b) shows.

\subsubsection{Generative Transformer and Masked MHSA Mechanism} The generative Transformer is one kind of autoregressive generative models that was proposed for text generation and is known as one of the most important driven forces for the transformative ChatGPT. The core of the generative Transformer is the masked MHSA mechanism, a pivotal component enabling the effective processing of input sequences. The masked MHSA mechanism allows the model to capture long-term dependencies among elements of input sequences and selectively attends to these elements while preventing it from looking ahead during training by masking future elements. To learn temporal dependencies efficiently for downlink channel extrapolation, we propose the generative Transformer-based downlink channel extrapolation network. The architecture shown in Fig. \ref{DCE} (c) depicts that the proposed generative Transformer consists of a temporal positional encoding layer and $N_G$ generative Transformer layers. The temporal positional encoding layer is similar to (\ref{PE1}) - (\ref{PE2}). Each generative Transformer layer further consists of a masked MHSA sub-layer, a feed-forward (FF) sub-layer, and also RC and LN operations around each of the two sub-layers. Compared with the MHSA sub-layer, a masking operation is introduced before the query, key, and value projections in the masked MHSA sub-layer. The output of the masking operation, denoted by $\tilde{\mathbf{X}}^M \in \mathbb{R}^{N_I^\prime \times d_R^\prime}$ with $N_I^\prime$ being the number of input elements and $d_R^\prime$ being the representation dimension of each input element, can be expressed by
\begin{equation}
\tilde{\mathbf{X}}^M = \tilde{\mathbf{X}}^\prime + \mathbf{M},
\end{equation}
where $\tilde{\mathbf{X}}^\prime \in \mathbb{R}^{N_I^\prime \times d_R^\prime}$ is the input matrix to the masked MHSA sub-layer, and $\mathbf{M} \in \mathbb{R}^{N_I^\prime \times d_R^\prime}$ is a lower-triangular masking matrix whose lower triangular elements are all negative infinity. Note that the masked MHSA sub-layer is with $N_h^\prime$ attention heads and a dropout probability $p_3$.
The FF sub-layer can be formulated by
\begin{equation}\label{FF}
\mathbf{X}^{FF} = (\operatorname{Drop}(\operatorname{ReLU}(\mathbf{X}^{RC1}\mathbf{W}^{FF1}+\mathbf{b}^{FF1});p_4))\mathbf{W}^{FF2}+\mathbf{b}^{FF2},
\end{equation}
where $\mathbf{X}^{FF} \in \mathbb{R}^{N_I^\prime \times d_R^\prime}$, $p_4$ is a dropout probability, $\mathbf{W}^{FF1} \in \mathbb{R}^{d_R^\prime \times d_{FF}}$ and $\mathbf{W}^{FF2} \in \mathbb{R}^{d_{FF} \times d_R^\prime}$ are learnable weight matrices, and $\mathbf{b}^{FF1} \in \mathbb{R}^{1 \times d_{FF}}$ and $\mathbf{b}^{FF2} \in \mathbb{R}^{1 \times d_R^\prime}$ are learnable bias vectors.
The two sets of RC and LN operations around the masked MHSA sub-layer and the FF sub-layer are organized and computed similarly to the two sets of RC and LN operations shown in (\ref{RC1_1}) to (\ref{LN1_3}) and (\ref{RC2_1}) to (\ref{LN2_3}), respectively.

\subsection{Computational Complexity Analysis}
Here we analyze the computational complexity of the spatial-frequency sampling embedding layer, and compare it with the most common existing spatial-frequency domain representation extractors for high-dimensional inputs: 1) the 2D convolution layer and 2) the fully-connected layer.

The time complexity of the spatial-frequency sampling embedding layer can be calculated as
\begin{equation} \label{sfse time}
    \mathcal{O}(\text{SFSE}) = \mathcal{O}(2 N_R N_T N_c d_R^\prime),
\end{equation}
and the space complexity (i.e., the number of learnable parameters) can be calculated as
\begin{equation} \label{sfse space}
    N_{\theta}(\text{SFSE}) = \mathcal{O}(2 N_R \frac{N_T}{N_1} \frac{N_c}{N_2} d_R^\prime).
\end{equation}

The 2D convolution layer is blamed for high time complexity and the fully-connected layer is blamed for high space complexity when dealing with high-dimensional inputs. The time complexity of the 2D convolution layer can be calculated as
\begin{equation} \label{conv time}
    \mathcal{O}(\text{2D Conv}) = \mathcal{O}(2 k^2 N_R N_T N_c d_R^\prime),
\end{equation}
where $k >= 1$ is the convolution kernel size. It is worth emphasizing that to reduce the spatial and frequency dimensions of the input, the stride of the 2D convolution layer should be set larger than $1$, or a pooling layer should be applied after the 2D convolution layer. Generally, it requires a deep convolutional network consisting of many 2D convolution layers to gradually reduce the spatial and frequency dimensions of the input. This thus further increases the time complexity of using 2D convolution layers for obtaining the spatial-frequency domain representation for high-dimensional CSI matrices. While for the spatial-frequency sampling embedding layer, one layer is enough to obtain the spatial-frequency domain representation for high-dimensional CSI matrices. Therefore, the spatial-frequency sampling embedding layer is $\gg k^2$ times more efficient than the deep convolutional network.

The fully-connected layer has the same time complexity as the spatial-frequency sampling embedding layer. However, the space complexity of the fully-connected layer is huge, which can be calculated as
\begin{equation} \label{FC space}
    N_{\theta}(\text{FC}) = \mathcal{O}(2 N_R N_T N_c d_R^\prime),
\end{equation}
which is $N_1 \times N_2$ times as the spatial-frequency sampling embedding layer, leading to prohibitive memory usage and a challenging optimization process when training the neural network.

\subsection{Model Training}
To train the model efficiently, the sliding window method is applied to generate windowed temporal channel slices. The window length is $l_w = N_{slot}$ and the sliding stride is $l_s = N_{slot}$. Denoting the number of training samples and the number of sub-frames as $N_{train}$ and $N_{sf}$, respectively, the normalized mean squared error (NMSE) loss function for training the UDCCN is defined as
\begin{equation}\label{L2}
    L_2 =  \mathbb{E}_{N_w}\left\{\frac{\|{\mathbf{H}^d} \{s,1\} - F_{udcc}(\hat{\mathbf{H}}{\{s,0\}})\|_2^2}{\|{\mathbf{H}^d} \{s,1\}\|_2^2}\right\},
\end{equation}
where $N_w = N_{train} * N_{sf}$ is the number of slices generated from all training samples. The NMSE loss function for training the DCEN is defined as
\begin{equation}\label{L3}
L_3 = \mathbb{E}_{N_w}\left\{\frac{\sum_{t=2}^{N_{slot}-1}\frac{\|{\mathbf{H}^d}\{s,t\} - F_{dce}(\hat{\mathbf{H}}^d\{s,1\})\|_2^2}{\|{\mathbf{H}^d}\{s,t\}\|_2^2}}{N_{slot}-2}\right\}.
\end{equation}
\section{Simulations} \label{Sec. Sim.}

In this section, we first introduce the simulation setup for numerical evaluations. Then, we compare the proposed KDD-SFCEN with traditional and DNN-based extrapolation methods under various spatial and frequency compression ratios, SNRs, and UE velocity settings. We then compare the TUDCEN with DNN-based temporal channel extrapolators. Finally, we present and analyze the overall performance of the proposed framework in terms of the achievable rate of the system.

\subsection{Simulation Setup}
\subsubsection{Simulation Dataset, Parameters, and Model Training}
The simulation dataset is constructed using the MATLAB 5G toolbox \cite{MATLAB5G}. We consider a mmWave massive MIMO system with one BS and one single user. Uniform linear arrays (ULAs) are employed at the BS and the UE with $N_T = 32$  and $N_R = 4$. The system works in the TDD mode and operates with the OFDM modulation. The system comprises $52$ RBs, and each RB is formed by $12$ subcarriers and $14$ OFDM symbols according to the 5G specification \cite{3gpp2023ts38211}. In addition, the system follows the 5G frame structure shown in Fig. \ref{system} (e) with $\mu=3$ and $N_{slot}=8$, and note that the SRS period $T_{p} = 1$ ms (i.e., once per sub-frame). The SRS is configured at the last symbol index of each special slot. We adopt the clustered delay line (CDL)-B MIMO channel model to generate channel instances\cite{3gpp2022tr38901} at various UE mobility settings. The CDL-B channel model uses multiple clusters to characterize multipath propagation and supports a frequency range from $0.5$ GHz to $100$ GHz, which is widely adopted for link-level evaluation. $95$ channel instances are generated for constructing training and validation samples, and the BS transmits the SRS to the UE via these channel instances at the special slot of each sub-frame in $10$ frames (i.e., in total $100$ sub-frames / $800$ slots) at a signal-to-noise ratio (SNR) level of $5$ dB (i.e., $\gamma_{SNR}=5$ dB) and the UE mobility of 60 km/h (i.e., $v=60$ km/h). $5$ channel instances are generated for constructing testing samples, and the BS transmits the SRS to the UE via these channel instances at the special slots of $100$ sub-frames at different SNR levels. The SRS pattern is configured based on the spatial compression ratio $R_s$ and the frequency compression level $R_f$ as depicted in Section \ref{Sec. Sys. C}. As a result, for a given ($R_s$, $R_f$) pair, $9,500$ samples are generated for training ($9,000$ samples) and validation ($500$ samples); for a given ($R_s$, $R_f$, $\gamma_{SNR}$, $v$) pair, $500$ samples are generated for testing. The spatial compression ratios, frequency compression ratios, SNR levels, UE mobility, and other key system parameters are listed in Table \ref{SimulationSetup}. The hyper-parameter settings of the proposed KDD-SFCEN and the TUDCEN are shown in Table \ref{Hyper1} and Table \ref{Hyper2}, respectively. The model training processes are as follows. The KDD-SFCEN for uplink channel estimation is first trained with $L_1$ in (37), followed by training the UDCCN for uplink-downlink calibration with $L_2$ in (51), and finally, the DCEN for downlink channel extrapolation is trained with $L_3$ in (52), all based on the supervised learning paradigm. Specifically, $\mathbf{H}{\{s,0\}}$, ${\mathbf{H}^d} \{s,1\}$, and $\mathbf{H}^d\{s,t\}$ for $t =2,3,\ldots,N_{slot}-1$ are provided as labels for training the three models, respectively. By minimizing the three loss functions on the training samples with the Adam optimizer \cite{kingma2014adam} and validating the models on the validation samples, all three models are effectively trained and validated.

\begin{table}[htbp]
\centering
\caption{System parameter settings.}
\label{SimulationSetup}
\resizebox{\columnwidth}{!}{%
\begin{tabular}{c|c}
\hline
\textbf{Parameter}            & \textbf{Value}                          \\ \hline
Channel model                 & CDL-B                                   \\ \hline
Carrier frequency             & $28$ GHz                                \\ \hline
SRS period $T_p$              & $1$ ms                                  \\ \hline
Subcarrier spacing            & $120$ kHz                               \\ \hline
The number of frames          & $10$                                    \\ \hline
Numerology $\mu$                         & $3$                          \\ \hline
The number of slots per sub-frame $N_{slot}$ & $8$                      \\ \hline
The number of sub-frames $N_{sf}$     & $100$                           \\ \hline
The number of BS antennas $N_T$                         & $32$                                    \\ \hline
The number of UE antennas $N_R$                         & $4$                                     \\ \hline
The number of subcarriers $N_c$                         & $624$                                   \\ \hline
Spatial compression ratio $R_s$                         & $1$, $2$, $4$, $8$                      \\ \hline
Frequency compression ratio $R_f$                         & $2$, $4$, $8$, $16$                     \\ \hline
SNR $\gamma_{SNR}$ (dB)          & $-5$, $0$, $5$ (training), $10$, $15$, $20$        \\ \hline
UE velocity $v$ (km/h)              & $5$, $15$, $30$, $60$ (training), $90$, $120$                               \\ \hline
The number of training $N_{train}$ & $9,000$                 \\ \hline
The number of validation samples $N_{valid}$ & $500$                 \\ \hline
The number of testing samples $N_{test}$ & $500$                                   \\ \hline
\end{tabular}
}
\end{table}

\begin{table}[htbp]
\centering
\caption{Hyper-parameter settings for the KDD-SFCEN.}
\label{Hyper1}
\begin{tabular}{c|c}
\hline
\textbf{Hyper-parameter}                        & \textbf{Setting} \\ \hline
{Representation dimension in the KDD-SFCEN $d_R$}     & $512$              \\ \hline
{No. of attention head $N_h$}                   & $4$                \\ \hline
{MHSA sub-layer dropout probability $p_1$}              & $0.5$                \\ \hline
{SEG sub-layer dropout probability $p_2$}              & $0.5$                \\ \hline
{Batch size}                        & $64$              \\ \hline
{Initial learning rate}             & $6e-5$             \\ \hline
\end{tabular}
\end{table}

\begin{table}[htbp]
\centering
\caption{Hyper-parameter settings for the TUDCEN.}
\label{Hyper2}
\begin{tabular}{c|c}
\hline
\textbf{Hyper-parameter}                        & \textbf{Setting} \\ \hline
{Calibration feature dimension   $d_f$}     & $32$              \\ \hline
{Convolution kernel size $k$}     & $3$              \\ \hline
{Embedding dimension $d_{emb}$}     & $512$              \\ \hline
{Spatial sampling factor $N_1$}       & $4$              \\ \hline
{Frequency sampling factor $N_2$}     & $12$              \\ \hline
{Representation dimension in the TUDCEN $d_R^\prime$}     & $512$              \\ \hline
{No. of attention head $N_h^\prime$}                   & $4$                \\ \hline
{MHSA sub-layer dropout probability $p_3$}              & $0.5$                \\ \hline
{FF sub-layer dropout probability $p_4$}              & $0.5$                \\ \hline
{No. of generative Transformer layers}                   & $4$                \\ \hline
{Batch size $N_b$}                        & $100$              \\ \hline
{Initial learning rate}             & $6e-5$             \\ \hline
\end{tabular}
\end{table}

\subsubsection{Evaluation Metrics}
For uplink channel estimation, uplink-downlink channel calibration, and also downlink channel extrapolation, the NMSE in dB is adopted to evaluate the channel estimation performance, which is defined as
\begin{equation}\label{NMSE}
\text{NMSE\{t\}} = 10 * \log \left(\frac{\sum_{j=1}^{N_{test}} \sum_{s=1}^{N_{sf}} \frac{\|\overline{\mathbf{H}} \{j,s,t\} - \hat{\overline{\mathbf{H}}} \{j,s,t\}\|_2^2}{\|\overline{\mathbf{H}} \{j,s,t\}\|_2^2}}{N_{test} N_{sf} }\right),
\end{equation}
where $N_{test}$ is the number of testing samples, and $\{j,s,t\}$ refers to the $t$-th slot of the $s$-th
sub-frame of the $j$-th testing sample. Specifically, for uplink channel estimation, $\overline{\mathbf{H}}$ refers to $\mathbf{H}$ and $t=0$; for  uplink-downlink channel calibration, $\overline{\mathbf{H}}$ refers to $\mathbf{H}^d$ and $t=1$; for downlink channel extrapolation, $\overline{\mathbf{H}}$ refers to $\mathbf{H}^d$, and $t=2,3,\ldots,N_{slot}-1$.

In addition, the achievable rate in bps/Hz is adopted to evaluate the system performance, which is defined as (\ref{rate}) at the top of the next page, where $\sigma_n^2=0.01$ is the noise power, and $\mathbf{F}^{RF}_i \in \mathbb{C}^{N_T \times N_{RF}}$ and $\mathbf{F}^{BB}_i \in \mathbb{C}^{N_{RF} \times N_R}$ are the analog precoding matrix and digital precoding matrix for the $i$-th subcarrier designed based on the estimated downlink CSI $\hat{\mathbf{H}}^d_i$, respectively.\footnote{Note that since the analog precoder can only control the phase of the signals and are typically frequency-independent, the same analog precoder will be applied to all subcarriers in practice when designing $\mathbf{F}^{RF}_i$.}

\begin{figure*}
\begin{equation}\label{rate}
R\{t\}= \frac{1}{N_{test} N_{sf} N_{c}} \sum_{j=1}^{N_{test}} \sum_{s=1}^{N_{sf}} \sum_{i=1}^{N_{c}} \log _2\left(\left|\mathbf{I}+\frac{\mathbf{H}^d_i \{j,s,t\} \mathbf{F}^{RF}_i \{j,s,t\} \mathbf{F}^{BB}_i \{j,s,t\} (\mathbf{F}^{BB}_i \{j,s,t\})^H (\mathbf{F}^{RF}_i \{j,s,t\})^H (\mathbf{H}^d_i \{j,s,t\})^H }{N_R \sigma_n^2}\right|\right).
\end{equation}
\end{figure*}

\subsection{Performance Evaluation of Proposed KDD-SFCEN}\label{Sec. Sim. B}

\begin{figure}[htbp]
\centering
\includegraphics[width=0.45\textwidth]{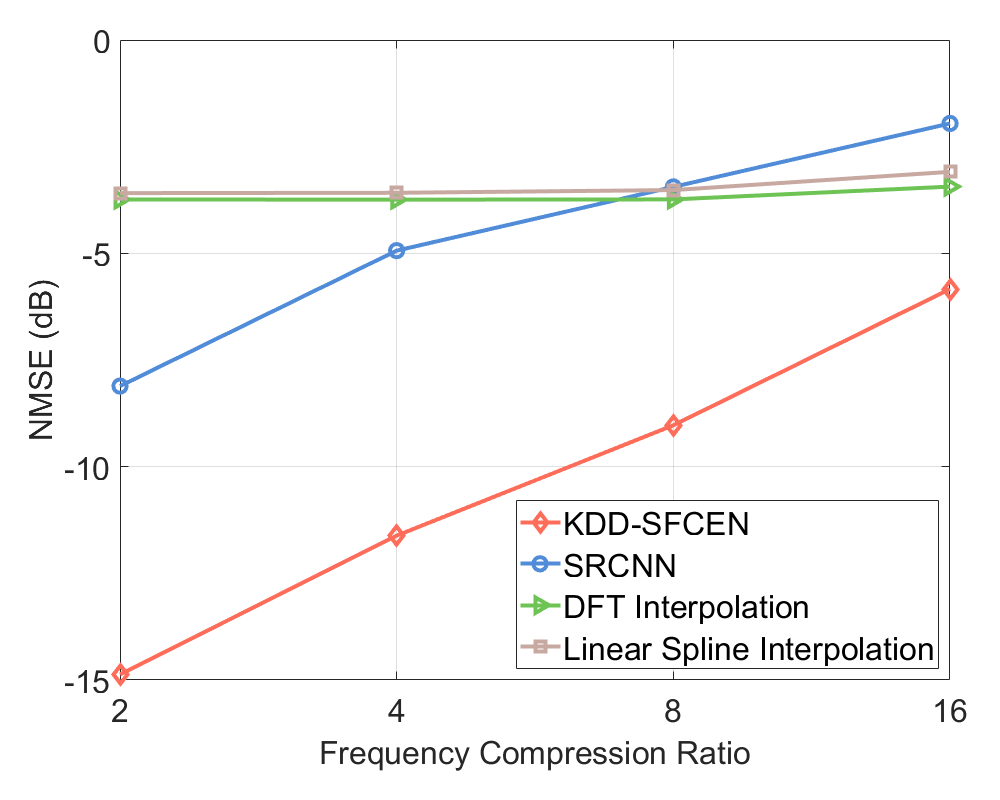}
\caption{The NMSE performance of uplink channel estimation versus the frequency compression ratio $R_f$ at $R_s=1$, $v=60$ km/h, and $\gamma_{SNR}=20$ dB.}
\label{FrequencyCR}
\end{figure}

We first compare the performance of the proposed KDD-SFCEN and traditional and DNN-based extrapolation methods for frequency domain channel extrapolation, including the linear spline interpolation method\cite{kusaykin2021based}, the DFT-interpolation method\cite{kusaykin2021based}, and the SRCNN-based method\cite{soltani2019deep}. It can be seen from Fig. \ref{FrequencyCR} that the proposed KDD-SFCEN outperforms all baselines at the same $R_f$. In addition, the proposed KDD-SFCEN can achieve a better NMSE performance at $R_f=16$ than traditional baselines at $R_f=2$, which indicates that the proposed KDD-SFCEN can reduce the pilot training overhead up to $8$ times from the frequency domain perspective compared with traditional baselines. Similarly, the proposed KDD-SFCEN can reduce the pilot training overhead up to $4$ times from the frequency domain perspective compared with the SRCNN-based method.

\begin{figure}[htbp]
\centering
\includegraphics[width=0.45\textwidth]{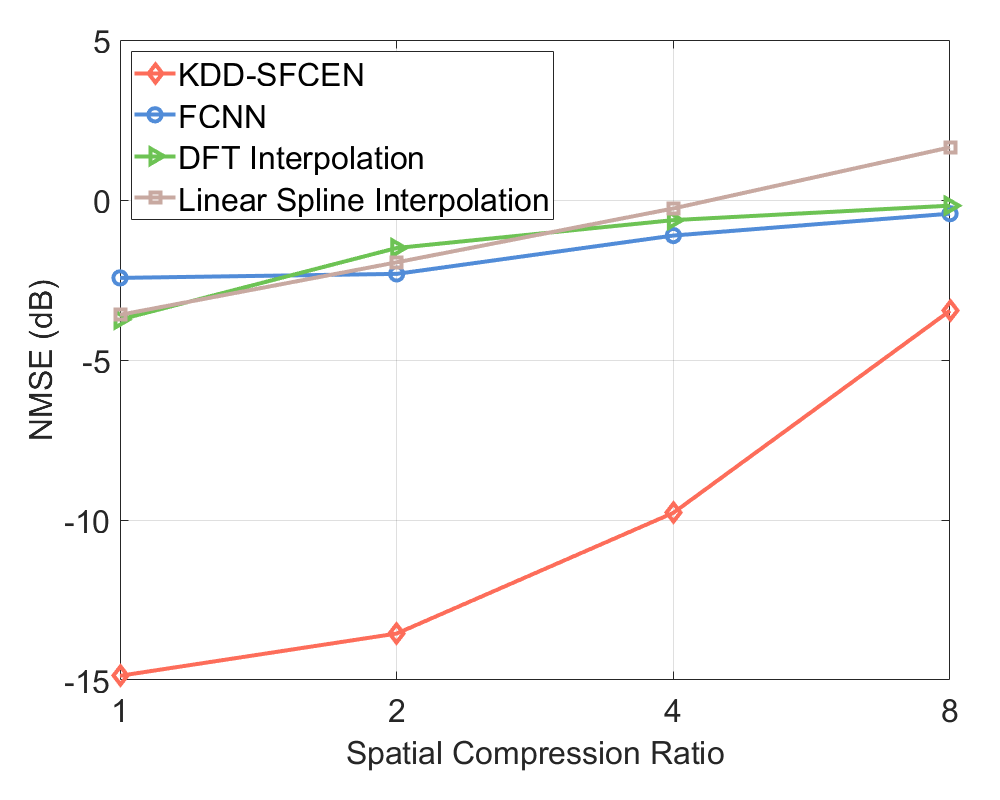}
\caption{The NMSE performance of uplink channel estimation versus the spatial compression ratio $R_s$ at $R_f=2$, $v=60$ km/h, and $\gamma_{SNR}=20$ dB.}
\label{SpatialCR}
\end{figure}

We then compare the proposed KDD-SFCEN with traditional and DNN-based extrapolation methods for spatial domain channel extrapolation, including the linear spline interpolation method\cite{kusaykin2021based}, the DFT-interpolation method\cite{kusaykin2021based}, and the FCNN-based method\cite{yang2020deepa}. As shown in Fig. \ref{SpatialCR}, the proposed KDD-SFCEN achieves better NMSE performance than all these baselines at the same $R_s$. Moreover, the proposed KDD-SFCEN at $R_s=8$ surpasses the FCNN-based method and achieves a similar NMSE performance as traditional baselines at $R_s=1$, indicating that the proposed KDD-SFCEN can reduce the pilot training overhead around $8$ times from the spatial domain perspective compared with existing baselines.

\begin{figure}[htbp]
\centering
\includegraphics[width=0.45\textwidth]{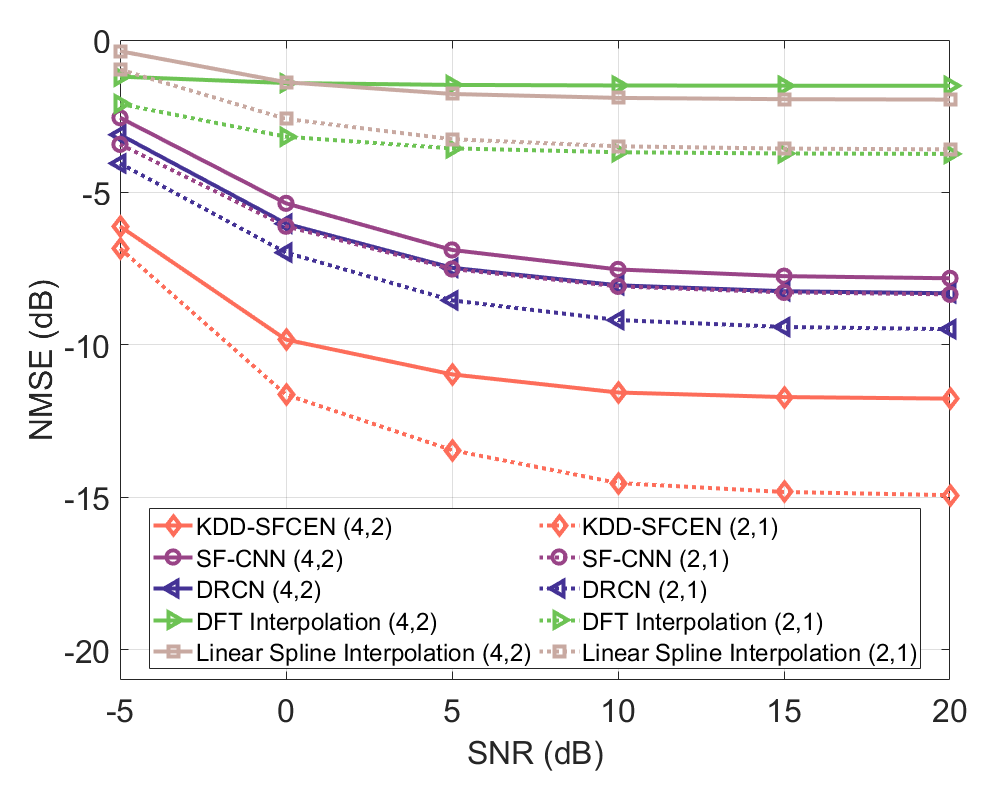}
\caption{The NMSE performance of uplink channel estimation versus the SNR $\gamma_{SNR}$ at $v=60$ km/h, with ($R_f=4$, $R_s=2$) or ($R_f=2$, $R_s=1$).}
\label{SFE_SNR}
\end{figure}

It can be seen from Fig. \ref{FrequencyCR} and Fig. \ref{SpatialCR} that when $R_f=8$ / $R_s=4$ (corresponding to a $4$-fold reduction in pilot training overhead compared to the minimum frequency / spatial compression ratio $R_f=2$ / $R_s=1$ according to Table \ref{SimulationSetup}), the NMSE performance of uplink channel estimation with the proposed KDD-SFCEN is around $-9$ dB / $-10$ dB at $\gamma_{SNR}=20$ dB, which is already good enough for uplink channel estimation. However, since the downlink CSI at future slots has to be extrapolated based on the estimated uplink CSI, it would be better to improve the performance of uplink channel estimation as much as possible. From Fig. \ref{FrequencyCR} and Fig. \ref{SpatialCR}, increasing $R_f$ from $2$ to $4$ or increasing $R_s$ from $1$ to $2$ leads to a minimum performance degradation compared to other increasing cases in $R_f$ or $R_s$.\footnote{Note that from Fig. \ref{FrequencyCR} that increasing $R_f$ leads to a near linear NMSE (in dB) performance degradation, thus the actual performance degradation increases in this progress.} These actually reflect the marginal effects of pilots in either the frequency domain or the spatial domain. Therefore, it is promising that we combine $R_f=4$ and $R_s=2$ for joint spatial and frequency domain channel extrapolation to improve the uplink channel estimation performance and restrict the performance degradation due to the reduction of pilot symbols in the frequency domain and the spatial domain. Fig. \ref{SFE_SNR} shows the NMSE performance of uplink channel estimation versus $\gamma_{SNR}$ at $R_f=4$ and $R_s=2$ (solid lines), which can also reduce $4$ times of pilot training overhead. The proposed KDD-SFCEN obtains an NMSE of $-12$ dB at $\gamma_{SNR}=20$ dB, indicating that joint spatial-frequency domain channel extrapolation can obtain a $3$ / $2$ dB of performance gain compared with individual frequency / spatial domain channel extrapolation. In addition, we further compare the proposed KDD-SFCEN with the DRCN\cite{liu2024learningbased}, the SF-CNN\cite{dong2019deep}, the linear spline interpolation method\cite{kusaykin2021based}, and the DFT-interpolation method\cite{kusaykin2021based} when $R_f=4$ and $R_s=2$ (solid lines) / $R_f=2$ and $R_s=1$ (dotted lines). The simulation results show that when $R_f=4$ and $R_s=2$, the KDD-SFCEN outperforms the DNN-based baselines (i.e., the DRCN and the SF-CNN) around $4$ dB and the traditional baselines around $9$ dB at $\gamma_{SNR} \ge 0$ dB; even with a low SNR, i.e., $\gamma_{SNR} = -5$ dB, the KDD-SFCEN can achieve a performance enhancement more than $3$ dB and $5$ dB compared with the DNN-based and traditional baselines, respectively. When $R_f=2$ and $R_s=1$, the KDD-SFCEN outperforms the DNN-based and traditional baselines more than $5$ dB and $11$ dB respectively at the high SNR regime, and about $3$ dB and $5$ dB respectively at the low SNR regime. Particularly, it can be found from Fig. \ref{SFE_SNR} that the NMSE performance of the proposed KDD-SFCEN with $R_f=4$ and $R_s=2$ is even better than existing methods with $R_f=2$ and $R_s=1$, indicating that the KDD-SFCEN can reduce the pilot training overhead up to $4$ times compared to existing methods.

\begin{figure}[htbp]
\centering
\includegraphics[width=0.45\textwidth]{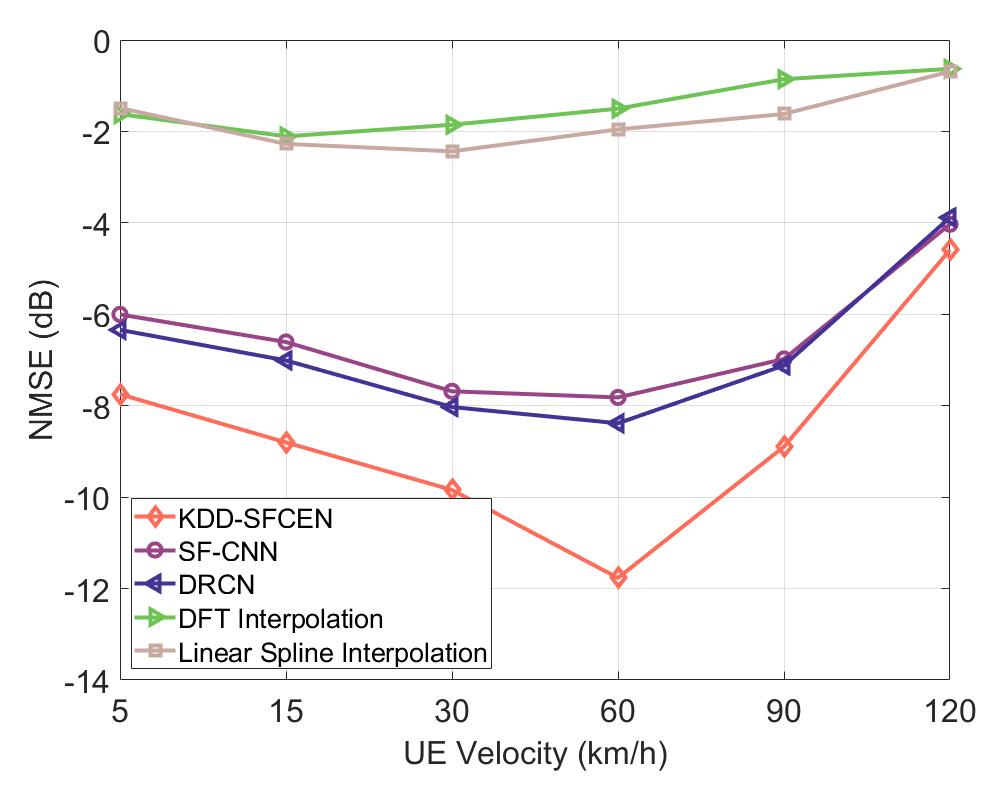}
\caption{The NMSE performance of uplink channel estimation versus the UE velocity $v$ at $R_f=4$, $R_s=2$, and $\gamma_{SNR}=20$ dB.}
\label{SFE_Velocity}
\end{figure}

We further examine the robustness of the proposed KDD-SFCEN to the UE velocity. As depicted in Fig. \ref{SFE_Velocity}, DNN-based methods (including the proposed KDD-SFCEN, the DRCN, and the SF-CNN) achieve the best performance when the testing UE velocity is equal to the training UE velocity (i.e., $v=60$ km/h), showing their capability in capturing spatial-frequency domain features and improving channel estimation accuracy. When the testing UE velocity decreases or increases, the performance of the DNN-based methods degrades. Moreover, the greater the deviation of the testing UE velocity from the training UE velocity, the more significant the performance degradation will be. Nonetheless, the proposed KDD-SFCEN achieves sufficient channel estimation accuracy ($\le -8$ dB) and outperforms all baselines with $v \le 90$ km/h at $R_f=4$ and $R_s=2$, showing its outstanding robustness to the UE velocity.

\subsection{Performance Evaluation of Proposed TUDCEN}
\begin{figure}[htbp]
\centering
\includegraphics[width=0.45\textwidth]{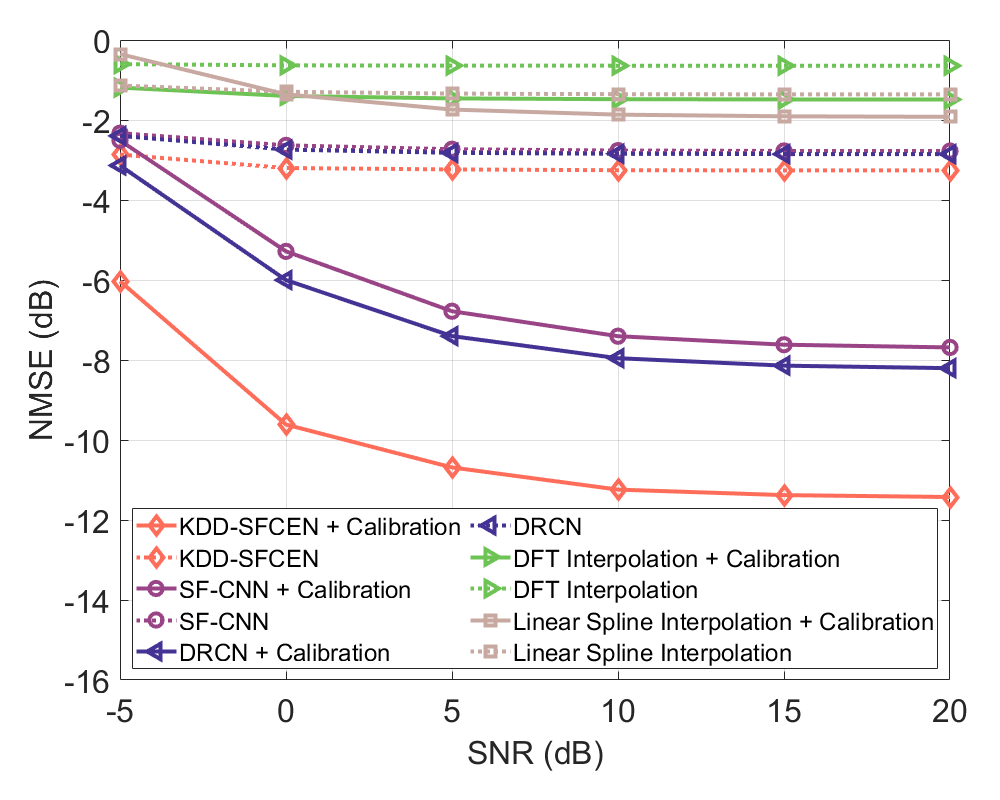}
\caption{The NMSE performance of uplink-downlink channel calibration versus the SNR $\gamma_{SNR}$ at $R_f=4$, $R_s=2$, and $v=60$ km/h.}
\label{TCE_Calibration_SNR}
\end{figure}

Since uplink-downlink channel calibration is simple but necessary, we do not compare the proposed UDCCN with other methods. Instead, we apply the UDCCN to the aforementioned spatial-frequency channel extrapolation methods and present their performance on estimating the channel at the first downlink slot with or without calibration. As shown in Fig. \ref{TCE_Calibration_SNR}, all methods generally benefit from the uplink-downlink channel calibration, demonstrating that uplink-downlink channel calibration is necessary and the proposed UDCCN is simple but effective to conduct uplink-downlink channel calibration.

\begin{figure}[htbp]
\centering
\includegraphics[width=0.45\textwidth]{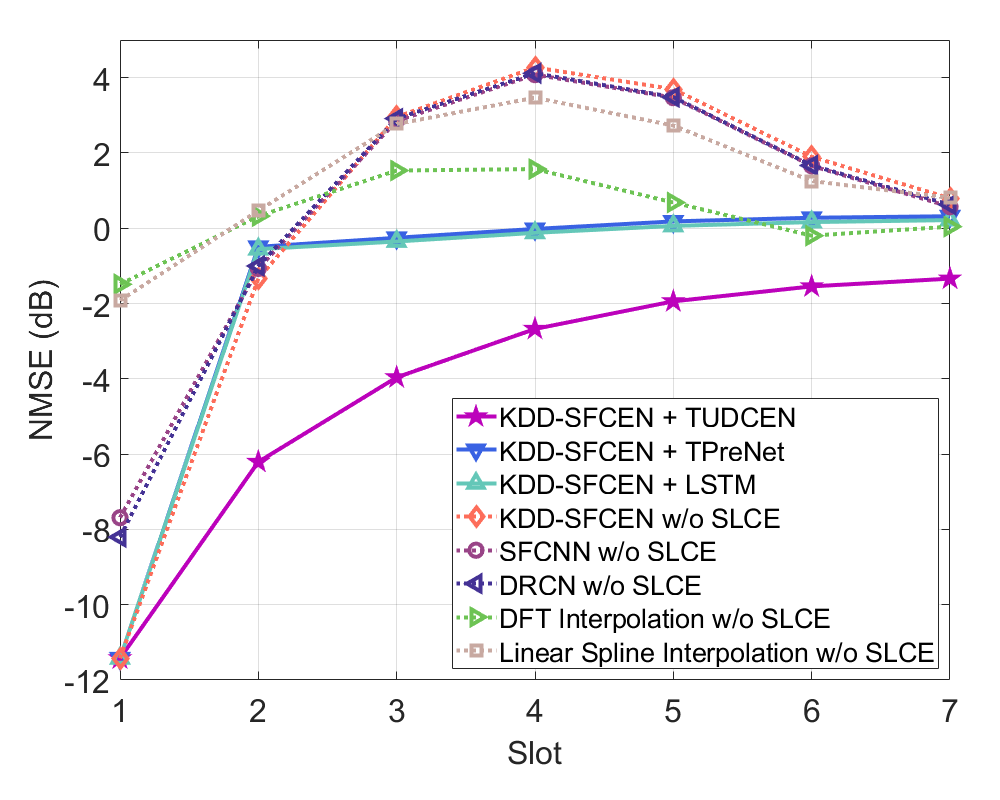}
\caption{The NMSE performance of slot-level channel extrapolation versus the index of a slot in a sub-frame at $R_f=4$, $R_s=2$, $\gamma_{SNR}=20$ dB, and $v=60$ km/h.}
\label{TCE_Step}
\end{figure}

\begin{figure}[htbp]
\centering
\includegraphics[width=0.45\textwidth]{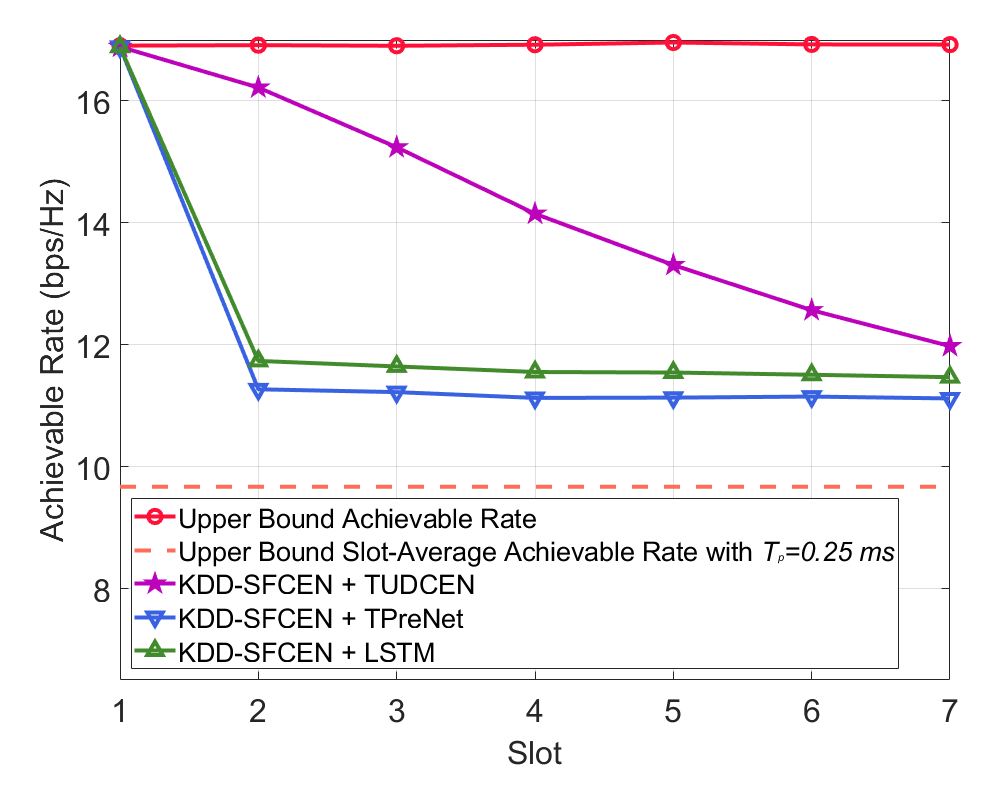}
\caption{The achievable rate performance of slot-level channel extrapolation versus the index of a slot in a sub-frame at $R_f=4$, $R_s=2$, $\gamma_{SNR}=20$ dB, and $v=60$ km/h.}
\label{TCE_Step_Rate}
\end{figure}

We then evaluate the performance of slot-level channel extrapolation with the proposed TUDCEN and two temporal channel extrapolation methods, i.e., the TPreNet\cite{gong2024lightweight} and the LSTM-based channel predictor\cite{mattu2022deep}. Note that the uplink-downlink channel calibration is applied to all methods. The NMSE performance and the achievable rate performance at $R_f=4$, $R_s=2$, $\gamma_{SNR}=20$ dB, and $v=60$ km/h from the first downlink slot to the seventh slot are shown in Fig. \ref{TCE_Step} and Fig. \ref{TCE_Step_Rate}, respectively. For better comparison, the upper bound achievable rate obtained with the perfect CSI and the upper bound slot-average achievable rate with a small SRS period $T_p=0.25$ ms (i.e., only $4$ slots are available for downlink transmission) are also shown in Fig. \ref{TCE_Step_Rate}.

It can be seen from Fig. \ref{TCE_Step} that without slot-level channel extrapolation (dotted lines), there are significant estimation errors for downlink channels at slots following the first downlink slot. Therefore, slot-level channel extrapolation is essential to reduce the channel estimation errors for these downlink channels. From Fig. \ref{TCE_Step} and Fig. \ref{TCE_Step_Rate}, compared with the proposed `KDD-SFCEN + TUDCEN', the performance of the `KDD-SFCEN + TPreNet' and `KDD-SFCEN + LSTM', in terms of either NMSE or achievable rate, degrades sharply to a poor regime at the second slot, indicating that these channel predictors are unable to deal with high-dimensional CSI matrices and provide accurate slot-level channel estimates. As depicted in Fig. \ref{TCE_Step_Rate}, the proposed `KDD-SFCEN + TUDCEN' can achieve near $80\%$ of the upper bound achievable rate even after $5$ slots from channel estimation. In addition, the achievable rate performance of the proposed `KDD-SFCEN + TUDCEN' is always better than the upper bound slot-average achievable rate with a small SRS period $T_p=0.25$ ms, demonstrating that with the proposed 3D channel extrapolation framework, the pilot signal can be sent less frequently (i.e., $T_p=1$ ms) while maintaining a satisfying achievable rate performance, thereby further reducing the pilot training overhead by $4$ times.

\section{Conclusion} \label{Sec. Con.}
This paper proposed a spatial, frequency, and temporal channel extrapolation framework for TDD mmWave massive MIMO-OFDM systems to systematically reduce the pilot training overhead under high-mobility scenarios. Specifically, two neural networks, namely the KDD-SFCEN and the TUDCEN, were proposed to reduce the spatial-frequency domain pilot training overhead and to reduce the temporal domain pilot training overhead, respectively. Extensive numerical results demonstrated that the proposed 3D channel extrapolation framework can effectively reduce the spatial-frequency domain pilot training overhead by $4$ times via spatial-frequency channel extrapolation and further reduce the temporal domain pilot training overhead by additional $4$ times via reducing the times of single-slot channel estimations with slot-level channel extrapolation.

\section{Acknowledgement}
This work was performed in part at SICC which is supported by SKL-IOTSC, University of Macau.

\bibliographystyle{IEEEtran}
\bibliography{refer}

\begin{IEEEbiography}
    [{\includegraphics[width=1in,height=1.25in,clip,keepaspectratio]{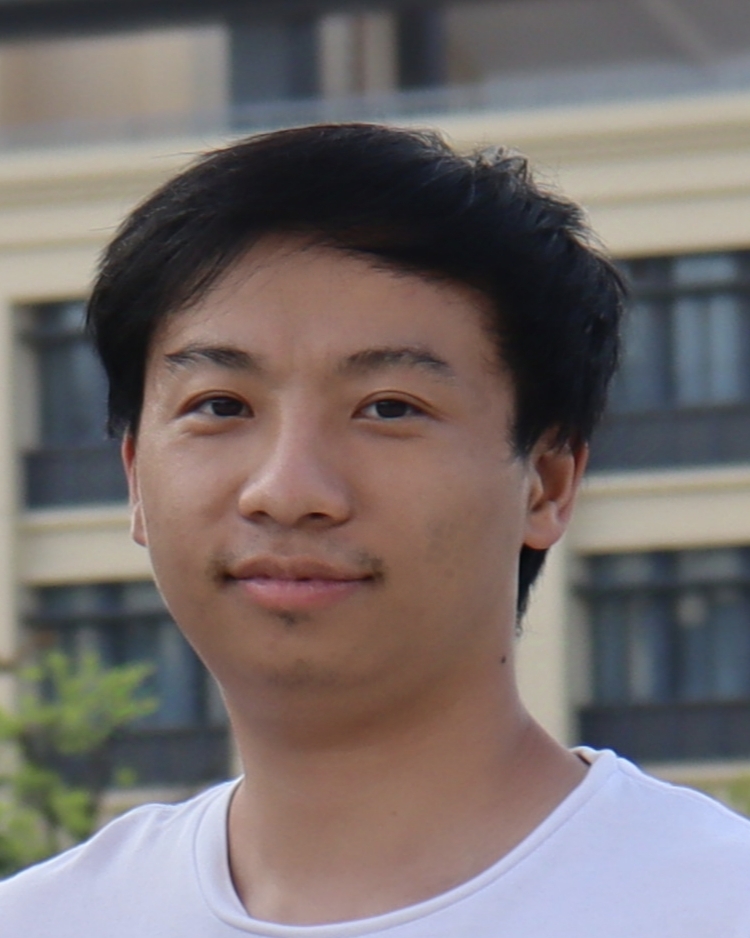}}]{Binggui Zhou} received his B.Eng. degree in Electrical Engineering from Jinan University, Zhuhai, China, in 2018, and his M.Sc. degree and Ph.D. degree in Electrical and Computer Engineering from the University of Macau, Macao SAR, China, in 2021 and 2024, respectively. He is currently a Postdoctoral Research Associate with the Department of Electrical and Electronic Engineering, Imperial College London, London, United Kingdom. He is also a Visiting Researcher with the School of Intelligent Systems Science and Engineering, Jinan University, Zhuhai, China. His research interests include machine learning, wireless communications, data mining, and smart healthcare.
\end{IEEEbiography}

\begin{IEEEbiography}
    [{\includegraphics[width=1in,height=1.25in,clip,keepaspectratio]{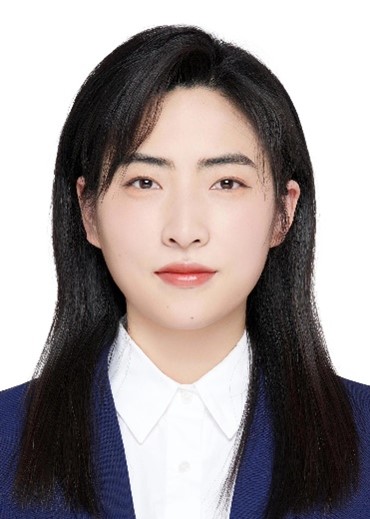}}]{{Xi Yang}} received the B.S., M.Eng. and Ph.D. degrees from Southeast University, Nanjing, China, in 2013, 2016 and 2019, respectively. From July 2020 to July 2022, she was a postdoctoral fellow with the State Key Laboratory of Internet of Things for Smart City, University of Macau, China. She is currently a Zijiang Young Scholar with the School of Communication and Electronic Engineering, East China Normal University, Shanghai, China. Her current research interests include extremely large aperture array (ELAA) massive MIMO communications, millimeter wave communications, and wireless communication system prototyping.
\end{IEEEbiography}

\begin{IEEEbiography}
    [{\includegraphics[width=1in,height=1.25in,clip,keepaspectratio]{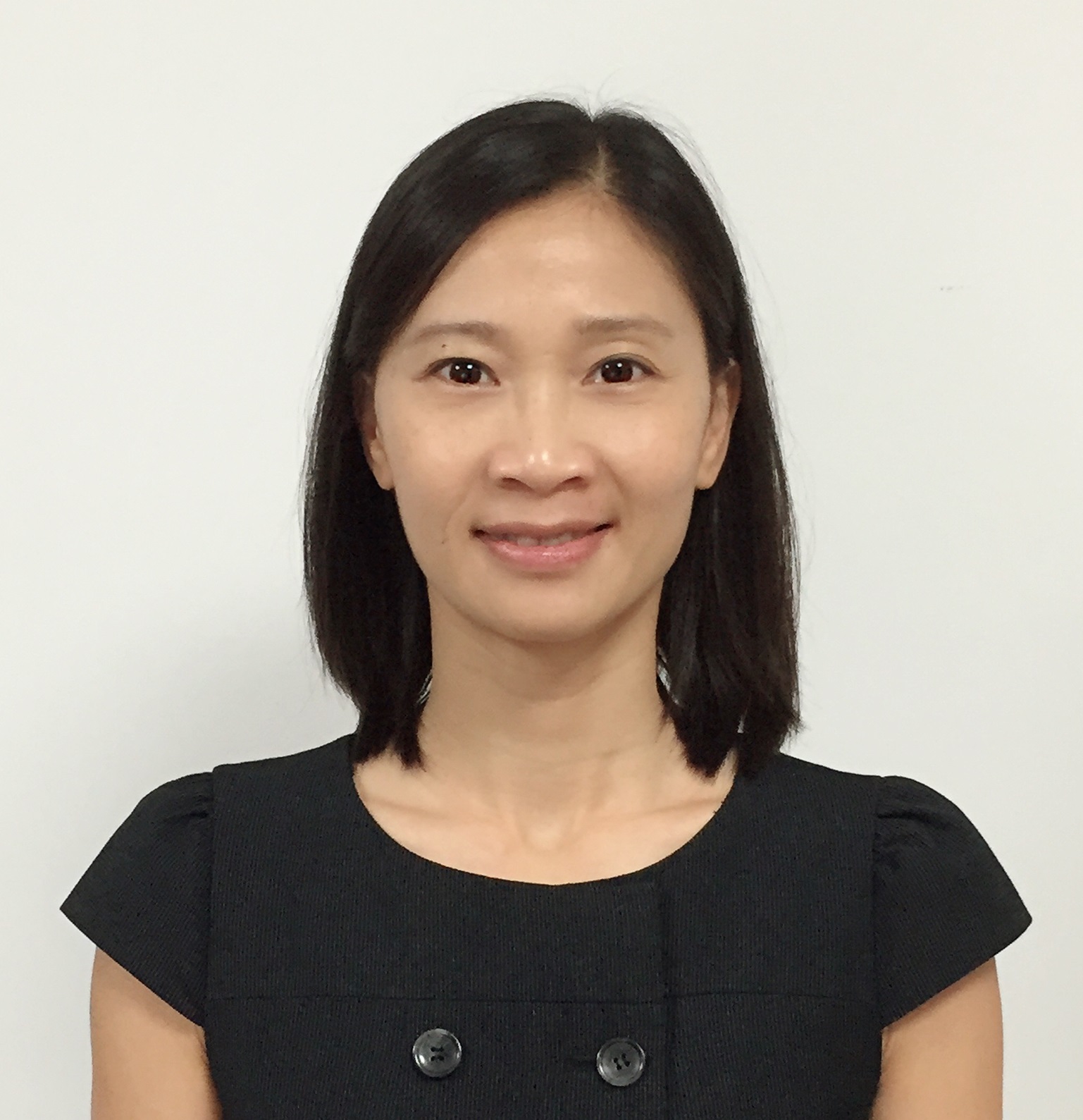}}]{Shaodan Ma} (Senior Member, IEEE) received the double Bachelor’s degrees in science and economics and the M.Eng. degree in electronic engineering from Nankai University, Tianjin, China, in 1999 and 2002, respectively, and the Ph.D. degree in electrical and electronic engineering from The University of Hong Kong, Hong Kong, in 2006. From 2006 to 2011, she was a post-doctoral fellow at The University of Hong Kong. Since August 2011, she has been with the University of Macau, where she is currently a Professor. Her research interests include array signal processing, transceiver design, localization, integrated sensing and communication, mmWave/THz communications, massive MIMO, and machine learning for communications. She was a symposium co-chair for various conferences including IEEE VTC2024-Spring, IEEE ICC 2021, 2019 \& 2016, IEEE GLOBECOM 2016, IEEE/CIC ICCC 2019, etc. She is an IEEE ComSoc Distinguished Lecturer in 2024-2025 and has served as an Editor for IEEE Wireless Communications (2024-present), IEEE Communications Letters (2023), Journal of Communications and Information Networks (2021-present), IEEE Transactions on Wireless Communications (2018-2023), IEEE Transactions on Communications (2018-2023), and IEEE Wireless Communications Letters (2017-2022).
\end{IEEEbiography}

\begin{IEEEbiography}
    [{\includegraphics[width=1in,height=1.25in,clip,keepaspectratio]{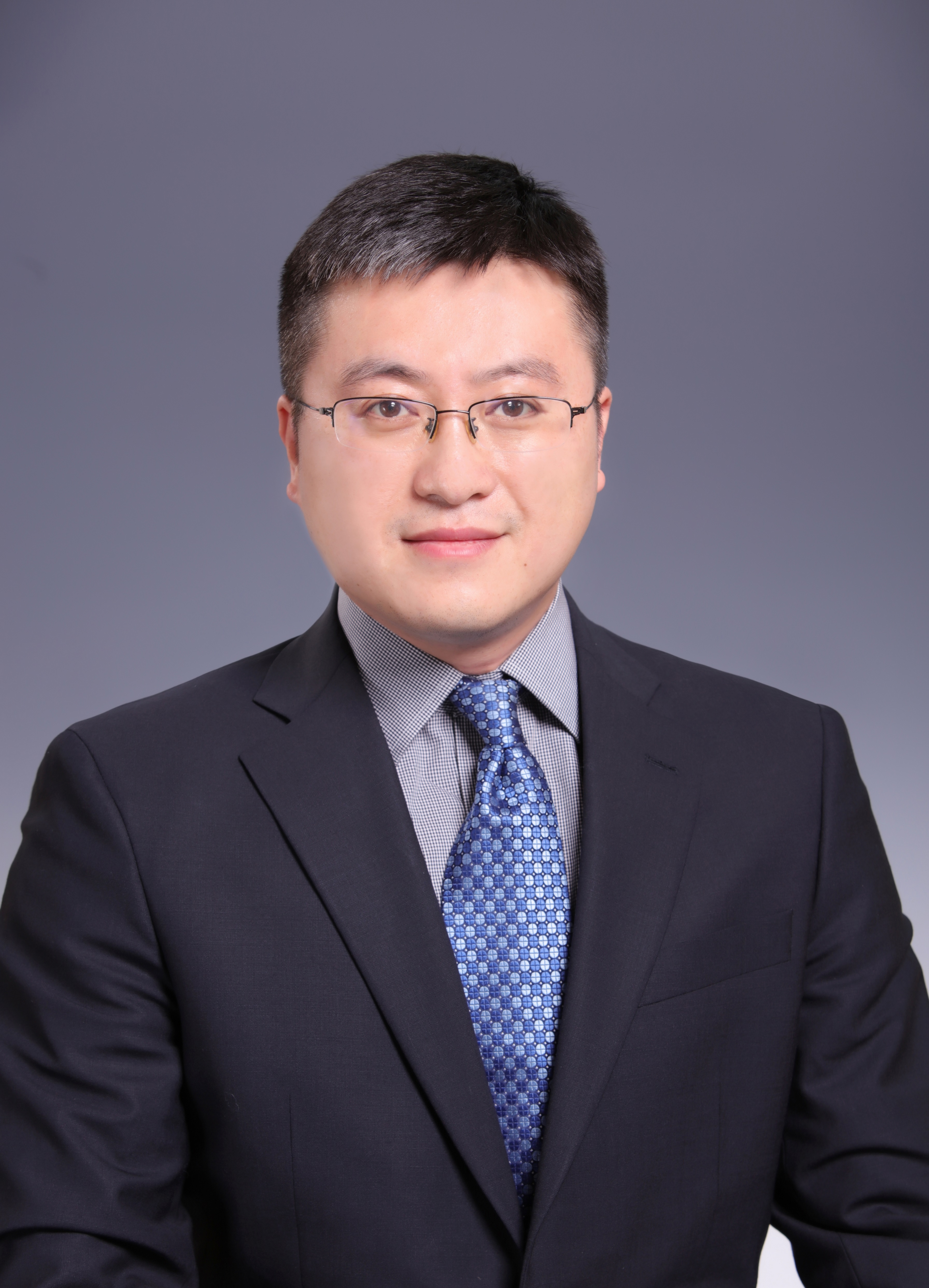}}] {Feifei Gao} (Fellow, IEEE) received the B.Eng. degree from Xi'an Jiaotong University, Xi'an, China in 2002, the M.Sc. degree from McMaster University, Hamilton, ON, Canada in 2004, and the Ph.D. degree from National University of Singapore, Singapore in 2007. Since 2011, he joined the Department of Automation, Tsinghua University, Beijing, China, where he is currently a tenured full professor. 

    Prof. Gao's research interests include signal processing for communications, array signal processing, convex optimizations, and artificial intelligence assisted communications. He has authored/coauthored more than 200 refereed IEEE journal papers and more than 150 IEEE conference proceeding papers that are cited more than 20000 times in Google Scholar. Prof. Gao has served as an Editor of IEEE Transactions on Wireless Communications, IEEE Journal of Selected Topics in Signal Processing (Lead Guest Editor), IEEE Transactions on Cognitive Communications and Networking, IEEE Signal Processing Letters (Senior Editor), IEEE Communications Letters (Senior Editor), IEEE Wireless Communications Letters, and China Communications. He has also served as the symposium co-chair for 2019 IEEE Conference on Communications (ICC), 2018 IEEE Vehicular Technology Conference Spring (VTC), 2015 IEEE Conference on Communications (ICC), 2014 IEEE Global Communications Conference (GLOBECOM), 2014 IEEE Vehicular Technology Conference Fall (VTC), as well as Technical Committee Members for more than 50 IEEE conferences.
\end{IEEEbiography} 

\begin{IEEEbiography}
    [{\includegraphics[width=1in,height=1.25in,clip,keepaspectratio]{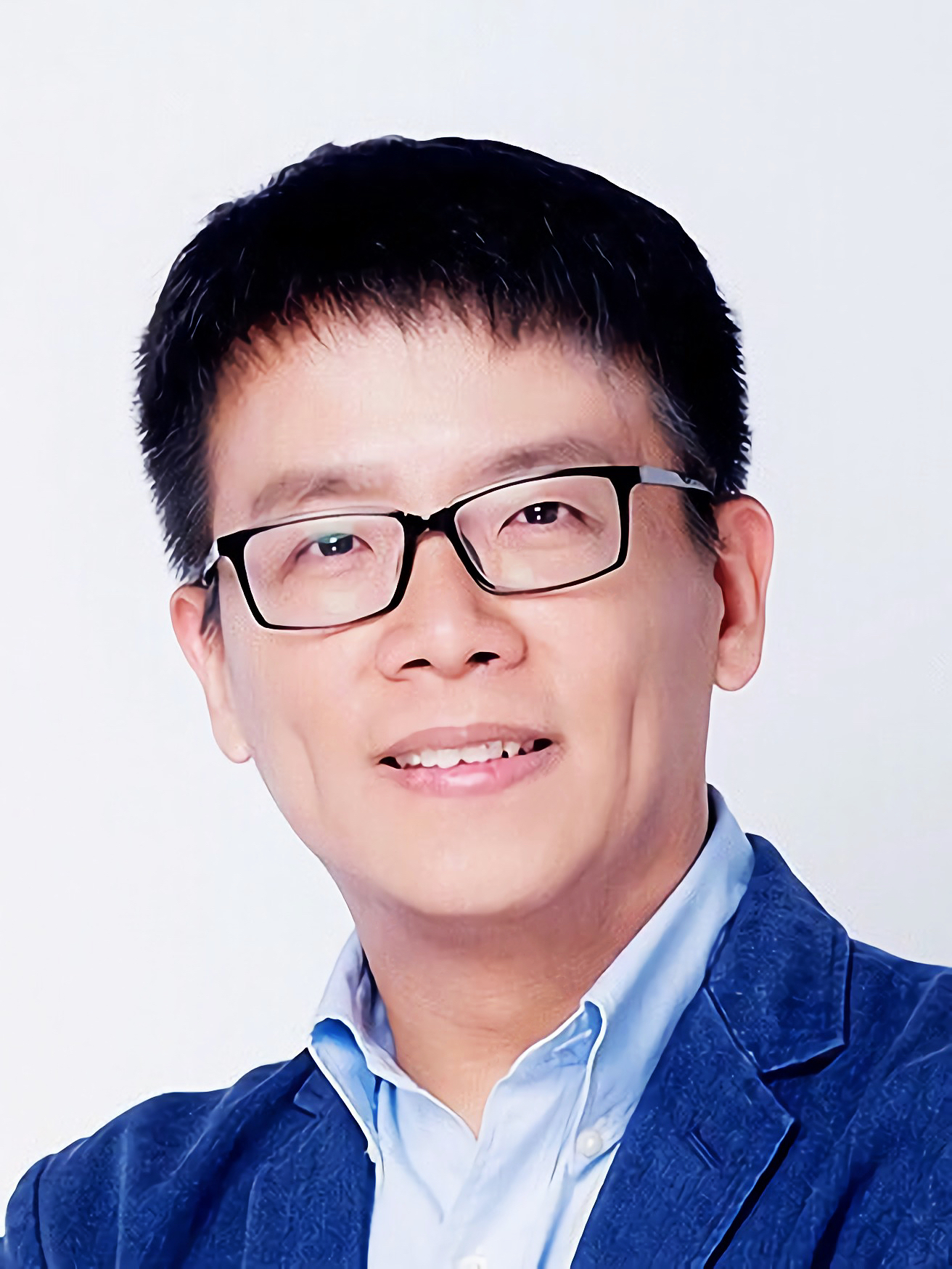}}]{Guanghua Yang} (Senior Member, IEEE) received his Ph.D. degree in electrical and electronic engineering from the University of Hong Kong in 2006. From 2006 to 2013, he served as post-doctoral fellow, research associate at the University of Hong Kong. Since April 2017, he has been with Jinan University, where he is currently a Full Professor in the School of Intelligent Systems Science and Engineering. His research interests are in the general areas of communications and networking.
\end{IEEEbiography}
\end{document}